\documentclass{emulateapj}

\usepackage{natbib}
\slugcomment{To Appear in The Astrophysical Journal}

\shorttitle{Arcturus}
\shortauthors{Ram\'irez and Allende Prieto}

\newcommand{\feh}{\mathrm{[Fe/H]}}
\newcommand{\teff}{T_\mathrm{eff}}
\newcommand{\logg}{\log g}

\newcommand{\fei}{Fe\,\textsc{i}}
\newcommand{\feii}{Fe\,\textsc{ii}}

\newcommand{\kms}{km\,s$^{-1}$}

\begin{document}

\title{Fundamental Parameters and Chemical Composition of Arcturus}

\author{I.\,Ram\'irez \altaffilmark{1} and
        C.\,Allende Prieto \altaffilmark{2,3}
	}

\altaffiltext{1}{The Observatories of the Carnegie Institution for Science,
                 813 Santa Barbara Street, Pasadena, CA 91101, USA}
\altaffiltext{2}{Instituto de Astrof\'isica de Canarias,
                 38205, La Laguna, Tenerife, Spain}
\altaffiltext{3}{Departamento de Astrof\'{\i}sica, Universidad de La Laguna,
                 38206, La Laguna, Tenerife, Spain}
\email{ivan@obs.carnegiescience.edu,callende@iac.es}

\begin{abstract}
We derive a self-consistent set of atmospheric parameters and abundances of 17 elements for the red giant star Arcturus: $\teff=4286\pm30$\,K, $\logg=1.66\pm0.05$, and $\feh=-0.52\pm0.04$. The effective temperature was determined using model atmosphere fits to the observed spectral energy distribution from the blue to the mid-infrared (0.44 to 10\,$\mu$m). The surface gravity was calculated using the trigonometric parallax of the star and stellar evolution models. A differential abundance analysis relative to the solar spectrum allowed us to derive iron abundances from equivalent width measurements of 37 \fei\ and 9 \feii\ lines, unblended in the spectra of both Arcturus and the Sun; the $\feh$ value adopted is derived from \fei\ lines. We also determine the mass, radius, and age of Arcturus: $M=1.08\pm0.06\,M_\odot$, $R=25.4\pm0.2\,R_\odot$, and $\tau=7.1^{+1.5}_{-1.2}$\,Gyr. Finally, abundances of the following elements are measured from an equivalent width analysis of atomic features: C, O, Na, Mg, Al, Si, K, Ca, Sc, Ti, V, Cr, Mn, Co, Ni, and Zn. We find the chemical composition of Arcturus typical of that of a local thick-disk star, consistent with its kinematics.
\end{abstract}


\section{Introduction}

The nearby K-giant Arcturus (HR\,5340, HD\,124897, HIP\,69673, $\alpha$ Boo) is an excellent reference for spectroscopic studies of giant stars. It is one of the brightest stars in the sky ($V=-0.05$\,mag) and its relatively low declination ($\delta\simeq+19^{\circ}$) makes it observable from most observatories in both hemispheres. Its moderately low metallicity ($\feh\simeq-0.5$) and Galactic space velocities associate the star with the Milky Way's thick disk \citep[e.g.,][]{ramirez07}. Moreover, the star has been identified as a member of a kinematic group. The so-called ``Arcturus group'' \citep{eggen71} stars have been proposed to be members of a dissolved stellar cluster or at least remnants of a dispersed short-lived star-forming event \citep{desilva07,williams09}. They have been even speculated to be of extragalactic origin \citep{navarro04}.

The atmospheric parameters of Arcturus have been estimated by several investigators using a variety of techniques. The PASTEL database of stellar parameters by \cite{soubiran10}, for example, lists 28 entries for Arcturus, with 24 of them including $\feh$ determinations from high resolution spectra. The simple mean and standard deviation for the stellar parameters compiled in PASTEL are: $\teff=4324\pm90$\,K, $\logg=1.71\pm0.29$, and $\feh=-0.56\pm0.10$. The published parameters do not distribute randomly around the mean values due to the impact of systematic errors which vary between different studies. Therefore, these average literature values are not precise. On the other hand, it is possible to obtain very precise atmospheric parameters for Arcturus using the best quality data and most reliable models available. Moreover, one can attempt to adopt the least model-dependent techniques to obtain not only precise but also accurate results.

Having precise and accurate atmospheric parameters for Arcturus is important for studies of giant stars in general. Since most systematic errors are dependent on the atmospheric parameters, differential analyses of giant stars relative to Arcturus can largely minimize those errors. A dramatic example of the power of differential analysis is provided by \cite{melendez09:twins} and \cite{ramirez09}, who performed strictly differential analyses of solar twin stars relative to the Sun to determine elemental abundances with unprecedented precision. The solar parameters are too different from those of the giant stars to be useful for extremely precise differential work. Arcturus, on the other hand, represents a much better reference for this purpose, and some studies have already taken advantage of this fact \citep[e.g.,][]{mcwilliam94,worley09,alves-brito10}.

Large spectroscopic surveys are producing enormous and homogeneous data sets. By properly analyzing them, our knowledge of galaxy formation and evolution will be greatly enhanced. We must realize, however, that large number statistics does not help removing systematic errors in the spectroscopic analysis, and it is therefore crucial to have reference stars with extremely well determined fundamental parameters and abundances. The Apache Point Observatory Galactic Evolution Experiment (APOGEE) survey, in particular, will observe about 100,000 giant stars, mainly in the Galactic bulge \citep{allende08:apogee,schiavon10,majewski10,shetrone10}. A precise and accurate determination of fundamental parameters and elemental abundances for the giant star Arcturus will therefore be of great importance for the proper handling of this large data set.

Arcturus has been analyzed many times in the past (see \citealt{mackle75} and \citealt{peterson93} for just two examples), but the availability of new data, including very high signal-to-noise high-resolution spectra, warrants a re-analysis. We use what we consider the most reliable methods and models to estimate Arcturus' atmospheric parameters and elemental abundances. Our work is based on classical static 1D-LTE model atmospheres and spectrum synthesis, but it sets a good starting point for future investigations on the impact of 3D and non-LTE effects \citep[e.g.,][]{asplund05:review}.

\section{Fundamental Parameters}

\subsection{Binarity}

Arcturus is flagged in the Hipparcos catalog as a two component object \citep{perryman97} but the adaptive optics observations by \cite{turner99} suggested that the star is single. Later, \cite{verhoelst05} found that a binary model with a G-type subgiant secondary matched their near-infrared interferometric data better than a single star model, implying that the Hipparcos flag is accurate. We find no detectable signatures of binarity in the high quality visible and near-infrared (normalized) spectrum of the star.

Arcturus has a chromosphere (see, e.g., \citealt{ayres75}) and that enhances visibly the continuum flux at wavelengths shorter than about 2000\,\AA. Compared to the Kurucz model atmosphere predictions for Arcturus, the star's absolutely calibrated fluxes in the ultraviolet, as given by \cite{ayres10}, reveal an excess at wavelengths between 2000 and 3000\,\AA\ that may not be related to the chromospheric activity of the star. This flux excess could be explained with a binary model in which the secondary is a slightly evolved warm star, such as that suggested by \cite{verhoelst05}. In this paper we study Arcturus as a single star because the impact of a possible secondary is only important in the ultraviolet, a wavelength region that we do not use and does not affect our analysis. We also note that there have been repeated reports in the literature about problems to understand the spectrum formed in the outermost layers of this star, in particular CO and H$_2$O transitions (e.g., \citealt{ayres86,ryde02,tsuji09}).

\subsection{Angular Diameter} \label{s:theta}

A number of interferometric measurements of the angular diameter of Arcturus have been published. They are listed in Table~\ref{t:angular-diameter} along with their sources. All reported measurements are consistent within the uncertainties. The angular diameter listed in the \cite{richichi09} paper, however, has a very small error compared to other sources, but it is likely underestimated, and the same, to less extent, is true for the result by \cite{perrin98}. This is most likely due to the fact that these reported values did not take systematics into account and are based on many observations, which reduces significantly the internal errors. According to \cite{mozurkewich03}, systematic uncertainties alone limit the precision of these measurements to no better than 1\,\%, which is about 0.2\,mas for Arcturus. Interestingly, this number is very similar to the average error bar published by the other groups listed in Table~\ref{t:angular-diameter}. Therefore, we assumed that the error bars for the \cite{richichi09} and \cite{perrin98} measurements are comparable to that from other sources, i.e., 0.2\,mas. Hereafter, the weighted mean and standard deviation from all the values given in Table~\ref{t:angular-diameter} is adopted as the angular diameter of Arcturus: $\theta_\mathrm{LD}=21.06\pm0.17$\,mas.

\begin{deluxetable}{lll}
\tablecolumns{3}
\tablewidth{0pc}
\tablecaption{Angular Diameter of Arcturus}
\tabletypesize{\footnotesize}
\tablehead{\colhead{$\theta_\mathrm{LD}$} & \colhead{error} & \colhead{Reference - Facility} \\ \colhead{(mas)} & \colhead{(mas)} & \colhead{}}
\startdata
20.95                & 0.20    & \cite{dibenedetto86} -- I2T \\
20.91                & 0.08    & \cite{perrin98} -- IOTA \\
21.0                 & 0.2     & \cite{quirrenbach96} -- MkIII \\
21.37                & 0.25    & \cite{mozurkewich03} -- MkIII \\ 
21.32                & 0.19    & \cite{verhoelst05} -- IOTA \\
21.05                & 0.21    & \cite{lacour08} -- IOTA \\
20.924               & 0.003   & \cite{richichi09} -- VLTI   \\ \hline
21.06                & 0.17    & Weighted mean (value adopted)\tablenotemark{1}
\enddata
\tablenotetext{1}{This value was obtained adopting a more conservative error of 0.20 mas for the \cite{perrin98} and \cite{richichi09} diameters.}
\label{t:angular-diameter}
\end{deluxetable}

Note that the values given in Table~\ref{t:angular-diameter}, and therefore the one adopted here, have been obtained from uniform disk measurements combined with limb-darkening corrections based on plane-parallel 1D-LTE model atmosphere predictions. Our adopted value for the angular diameter is thus consistent with the rest of our work, which is entirely based on the same type of atmospheric models.

\begin{figure}
\includegraphics[bb=95 370 560 700,width=9.0cm]{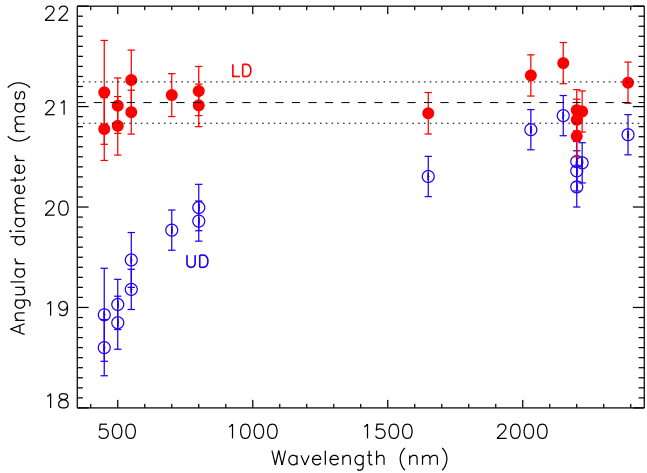}
\caption{Angular diameter measurements as a function of wavelength. Uniform-disk measurements are shown with open circles (sources are listed in Table~\ref{t:angular-diameter}). Limb-darkened diameters, obtained using the corrections by \cite{davis00}, are shown with filled circles. The dashed line shows the weighted average and the dotted lines are the $\pm1\,\sigma$ limits}
\label{f:theta}
\end{figure}

The limb-darkening corrections adopted by each study listed in Table~\ref{t:angular-diameter} are not all from the same source. To investigate whether this inconsistency is adding scatter to the average value that we adopt, we applied the limb-darkening corrections computed by \cite{davis00} using Kurucz model atmospheres to all of the uniform-disk measurements found in the literature. We used the \citeauthor{davis00} corrections as a function of wavelength for $\teff=4250$\,K, $\logg=1.5$, and $\feh=-0.5$, which correspond to the node of their grid that is nearest to the stellar parameters that we derive for Arcturus. As before, we adopted a minimum error bar of 0.2\,mas for the published diameters. We obtained $\theta_\mathrm{LD}=21.04\pm0.20$\,mas (see Fig.~\ref{f:theta}), which is consistent with the value we adopted but has a larger scatter. Thus, averaging true angular diameters computed with different limb-darkening corrections does not introduce scatter compared to the case when uniform-disk measurements from different sources are corrected for limb-darkening in a consistent manner.

Although, based on what can be found in the literature, the angular diameter of Arcturus appears to be known with high precision (better than 1\,\%\ after averaging independent measurements), we should remark that these results are model-dependent. For example, the use of spherically symmetric atmospheres instead of plane-parallel models results in an increase of $\sim0.5$\,mas for the angular diameter according to \cite{verhoelst05}. Moreover, the atmospheres of evolved stars are dynamically unstable and their surfaces are not spherical. This will have an impact on the measurement of uniform-disk diameters, which assume symmetric shapes, and limb-darkening corrections which are based on model atmosphere calculations \citep[see, e.g.,][]{koesterke08,chiavassa10}. In fact, 3D corrections to the limb-darkening may compensate the increase of angular diameter suggested by static spherical models \cite[e.g.,][]{allende02}. This possible systematic error will propagate in our analysis but not change our results significantly, as shown later. Unless otherwise noted, the angular diameter adopted in this work is that given in Table~\ref{t:angular-diameter}.

\subsection{Effective Temperature} \label{s:teff}

The shape of the spectral energy distribution (SED) of a star is determined primarily by its effective temperature. Certain features are also sensitive to other stellar parameters such as $\logg$ (e.g., the Balmer jump) or $\feh$ (e.g., the G-band and Ca~\textsc{ii} H\&K regions). If the latter can be determined accurately using independent methods, however, model fits to the observed SED can be used to estimate $\teff$ with high precision \citep[e.g.,][]{ramirez06}.

We use least-squares minimization to estimate the effective temperature of Arcturus from model fits to spectrophotometric data in the visible, near-infrared, and near- to mid-infrared, as described in detail below. The theoretical fluxes employed are from the Kurucz grid of no-overshoot model atmospheres with $\alpha$-element enhanced composition ($[\alpha/\mathrm{Fe}]=+0.4$; e.g., \citealt{castelli03}).\footnote{Available online at http://kurucz.harvard.edu/grids.html. These model fluxes have a spectral resolution of $R=\lambda/\Delta\lambda=100-500$ and wavelength sampling with steps of 20\,\AA\ in the visible and 50--200\,\AA\ at $\lambda>1\,\mu$m.} The use of plane-parallel model atmospheres for Arcturus is well justified here \citep{verhoelst05}, although sphericity could have an impact on the derivation of the angular diameter from interferometric measurements. The observed SEDs are divided by the scale factor $s=\theta_\mathrm{LD}^2/4$ so that they represent the flux that emerges from the surface of the star. For each data set, the spectra (observed or theoretical) were smoothed to a common spectral resolution (the one that corresponds to the lower resolution spectrum). A surface gravity $\logg=1.66$ (Sect.~\ref{s:logg}) and iron abundance $\feh=-0.52$ (Sect.~\ref{s:feh}) were adopted for these fits.

As usual, if the grid of model atmosphere fluxes is given by $\psi_\lambda(\teff)$, the reduced $\chi^2$ value is:
\begin{equation}
\chi_\nu^2 = \frac{1}{n-2} \sum_{i=1}^n \frac{(f_\lambda-\psi_\lambda)^2}{(\Delta f_\lambda)^2}\ ,
\end{equation}
where $f_\lambda$ are the observed fluxes, $\Delta f_\lambda$ their errors, and $n$ the number of observational data points. If available, we used the published $\Delta f_\lambda$ values, otherwise we adopted a constant 2\,\% error. The minimization of a set of $\chi_\nu^2$ values allowed us to obtain the best-fit $\teff$. We computed $\chi_\nu^2$ values in steps of 10\,K from $\teff=4220$ to 4350\,K for the visible fits (Sect.~\ref{s:sed_visible}) and in steps of 50\,K from $\teff=3900$ to 4600\,K for the infrared fits (Sects.~\ref{s:sed_nir} and \ref{s:sed_nmir}). To obtain a more precise effective temperature, we fitted a parabola to the 7 points nearest the minimum $\chi_\nu^2$.

Our least-squares minimization scheme uses only $\teff$ as a free parameter. Surface gravity and iron abundance are kept constant. The error in our $\teff$ estimate is thus given by:
\begin{equation}
\Delta\teff=\left( \frac{2}{\partial^2\chi_\nu^2/\partial\teff^2} \right)^{1/2} + \left| \frac{\partial\teff}{\partial s}\right|\Delta s \nonumber \ ,
\end{equation}
where the second term on the right-hand side of this equation corresponds to the error introduced by the uncertainty in the scale factor $s=\theta_\mathrm{LD}^2/4$ and therefore depends on the precision of the angular diameter measurement. The errors introduced by the uncertainties in $\logg$ and $\feh$ are very small compared to the other sources of error. Independently of the input data set, we find $\Delta\teff\lesssim\pm1$\,K using our derived values of $\Delta\logg=\pm0.06$ and $\Delta\feh=\pm0.04$. Nevertheless, this small error was added linearly to each $\Delta\teff$ estimate. Note that because we consider $s$, $\logg$, and $\feh$ fixed in the calculation of $\chi_\nu^2$, their contributions to $\Delta\teff$ are added linearly; the $\chi_\nu^2$ values depend on the adopted values of these fixed parameters.

We note that the Kurucz theoretical fluxes used here are not the result of radiative transfer calculations with a frequency step fine enough to guarantee that all spectral lines are well sampled. Nevertheless, we performed tests that show that this will not lead to a significant offset in the derived effective temperature.

\subsubsection{Visible} \label{s:sed_visible}

\begin{figure}
\includegraphics[width=8.8cm,bb=60 350 590 975]{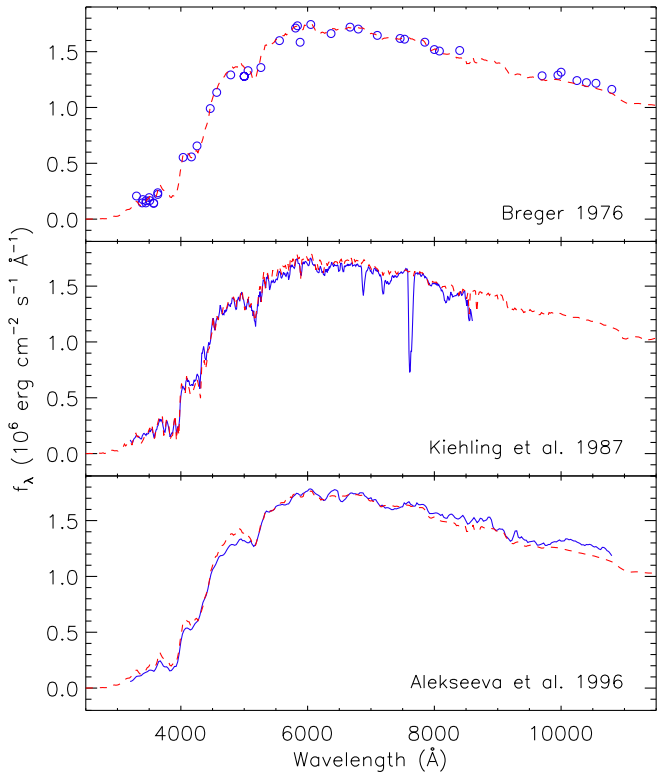}
\caption{Spectral energy distribution of Arcturus, scaled to represent the monochromatic flux that emerges from its surface. Open circles and solid lines represent observed data (references are given in the lower right of each panel). Dotted lines correspond to the best fit model atmosphere fluxes.}
\label{f:sed_vis}
\end{figure}

We use three published observed visible SEDs: \cite{breger76}, \cite{kiehling87}, and \cite{alekseeva96}. The Kurucz model fluxes have a finer sampling than the \cite{breger76} and \cite{alekseeva96} data so the resolution of the theoretical spectra was degraded when obtaining $\teff$ from those data sets. The opposite was the case for \cite{kiehling87}. The data and best model fits are shown in Fig.~\ref{f:sed_vis}. The \cite{kiehling87} data are severely affected by strong atmospheric absorption features such as O$_2$ and H$_2$O, which are very important at wavelengths longer than about 6800\,\AA. We excluded these regions from the model fits. At shorter wavelengths the \cite{alekseeva96} data appear to underestimate the UV-blue fluxes. Note, however, that in several other regions the fluxes from this source also show important discrepancies with the best model predictions. Both the \cite{breger76} and \cite{alekseeva96} fluxes appear too high relative to the best fit models near $\lambda=1\,\mu$m. To ensure consistency in the model fits to these three visible data sets, we used a single sample region from 4400 to 6800\,\AA. The outcome of our calculations does not change significantly if wider regions are used, but the model fits do not perform as well as in this wavelength window.

\begin{figure}
\includegraphics[bb=60 340 480 590,width=8.8cm]{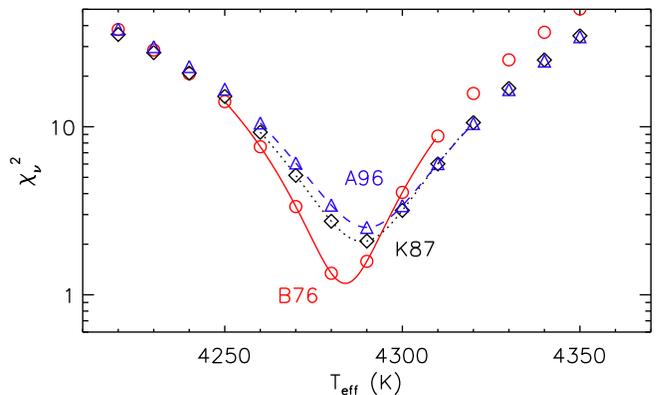}
\caption{Reduced $\chi^2$ value as a function of model $\teff$ for the three visible data sets. The exact wavelength range used in these computations is 4400--6800\,\AA. Solid, dotted, and dashed lines correspond to parabolic fits of the 7 points closest to the minima.}
\label{f:sed_vis_chi2}
\end{figure}

The run of $\chi_\nu^2$ values as a function of $\teff$ for the three observed fluxes is shown in Fig.~\ref{f:sed_vis_chi2}. The best fit $\teff$ values for the three observed data sets are consistent within a few degrees. Thus, we adopt a weighted average value as representative of the $\teff$ that corresponds to the SED in the visible: $\teff(\mathrm{visible})=4288\pm17$\,K.

\subsubsection{Near-IR (1--4 $\mu$m)} \label{s:sed_nir}

The spectral energy distribution of Arcturus in the near infrared, as measured by NASA's Infrared Telescope Facility (IRTF), is given by \cite{rayner09}. These data cover the region from about 1 to 4\,$\mu$m, excluding two relatively small windows around 1.9 and 2.7\,$\mu$m, which were discarded due to strong telluric absorption (cf.\ Fig.~3 in \citealt{rayner09}). We degraded the spectral resolution of these data ($2000\leq R\leq2500$) to match that of the Kurucz model fluxes. The \cite{rayner09} spectrum has been absolutely flux calibrated using 2MASS photometry \citep{skrutskie06}. Note, however, that 2MASS photometry is heavily saturated in the case of Arcturus and the value listed in the catalog was obtained by fitting the wings of the point-spread-function instead of the full PSF, as was the case for most other fainter stars. The uncertainties in the 2MASS photometry could be underestimated for this object. Thus, we introduced a second free parameter in our least-squares fits, namely a constant factor to multiply the fluxes, in order to take into account a possible large systematic error introduced by the uncertain 2MASS photometry.

\begin{figure}
\includegraphics[bb=60 340 480 590,width=8.7cm]{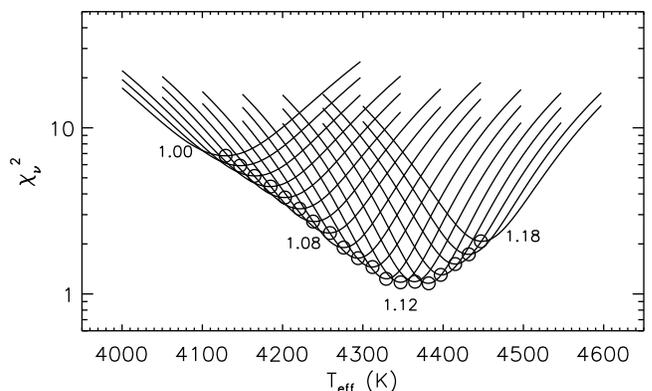}
\caption{Reduced $\chi^2$ value as a function of model $\teff$ for the IRTF data set multiplied by different flux scale factors from 1.00 to 1.18. The minimum in each case is shown with an open circle.}
\label{f:sed_irtf_chi2}
\end{figure}

In Fig.~\ref{f:sed_irtf_chi2} we show the $\chi_\nu^2$ values as a function of model $\teff$ for flux scale factors from 1.00 to 1.18. Clearly, a better fit is obtained after the published fluxes are multiplied by 1.12. The best fit model in this case has $\teff(\mathrm{IRTF})=4347\pm69$\,K. The fact that the published fluxes need to be scaled up by 12\,\%\ in order to be more consistent with the models and the observed visible SED suggests that the published 2MASS photometry for Arcturus is indeed uncertain and should be avoided.

\begin{figure}
\includegraphics[bb=60 360 480 1032,width=7.7cm]{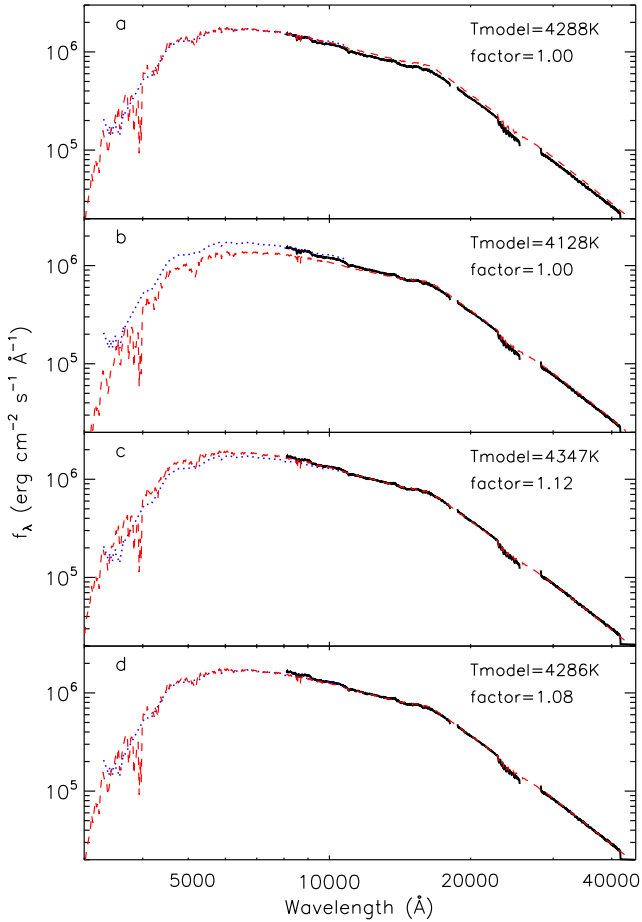}
\caption{Spectral energy distribution of Arcturus from the blue to the near infrared. Solid lines correspond to the IRTF data. Dotted lines are the visible SED by \cite{breger76}. Dashed lines correspond to the model atmosphere predictions ($\teff$ is given in the upper right corner of each panel). In each panel the IRTF fluxes have been scaled by the factor shown in the upper right corner.}
\label{f:sed_irtf}
\end{figure}

The IRTF fluxes are shown in Fig.~\ref{f:sed_irtf} for various values of the empirical flux scale factor. It is clear that the published fluxes (un-scaled) are too low compared to a model with the $\teff$ that corresponds to the visible fit (Fig.~\ref{f:sed_irtf}a). The best fit $\teff$ for this case ($=4128$\,K) is low compared to that obtained with the visible SEDs and severely underestimates the fluxes there (Fig.~\ref{f:sed_irtf}b). Our best fit model (factor=1.12 and $\teff=4347$\,K) does a much better job in reproducing the fluxes from about 1.5 to 4\,$\mu$m but overestimates the visible fluxes (Fig.~\ref{f:sed_irtf}c). We also note that a factor of 1.08 corresponds to a $\teff$ very close to that derived from the visible SEDs. However, the visible and near infrared SEDs do not connect smoothly around 1\,$\mu$m (Fig.~\ref{f:sed_irtf}d). Excluding the region from 0.8 to $1.5\,\mu$m, a $\teff=4286$\,K fits reasonably well both the visible and the near infrared data. Hereafter, whenever they are used, the IRTF fluxes are multiplied by 1.08. However, we stress the fact that the best model fit to the IRTF data alone is obtained with a scale factor of 1.12 and $\teff(\mathrm{IRTF})=4347\pm69$\,K, and this is the value adopted when averaging out the temperatures derived from different spectral windows.

\subsubsection{Near- to Mid-IR (2--10 $\mu$m)} \label{s:sed_nmir}

\begin{figure}
\centering
\includegraphics[bb=80 365 540 704,width=7.9cm]{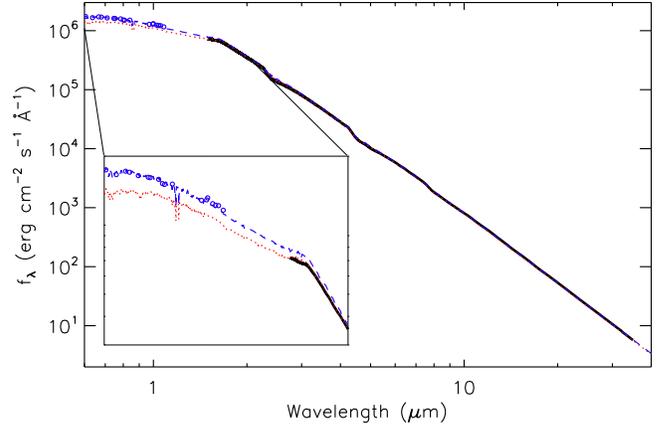}
\caption{Spectral energy distribution of Arcturus from the blue to the mid-infrared. Visible data (open circles) are from \cite{breger76}, near infrared data (dashed line) are from IRTF, and the near- to mid-infrared data (thick solid line) are from SWS. The dotted line corresponds to the best fit model. The bottom left corner inset shows an expanded view of the 0.6 to 2.0\,$\mu$m region.}
\label{f:sed_sws}
\end{figure}

In this spectral region, absolutely calibrated fluxes are available from \cite{engelke06}. These data were smoothed to match the spectral resolution of the theoretical fluxes. The \cite{engelke06} SED is based on data taken with the Short Wavelength Spectrometer (SWS) on the Infrared Space Observatory. Fig.~\ref{f:sed_sws} shows the SWS data along with visible and near-IR fluxes. The theoretical fluxes have a good wavelength coverage up to 10~$\mu$m but it becomes very sparse for longer wavelengths. In fact, between 10 and 40\,$\mu$m, only two theoretical data points are available for comparison with the observed data. Our least-squares minimization was restricted to the range from 2 to 10~$\mu$m. Visual inspection reveals that the two model flux points beyond 10~$\mu$m are fully consistent with the best model obtained from the shorter wavelength data. The best fit model was obtained, as usual, with the least-squares technique, and found to be $\teff(\mathrm{SWS})=4152\pm84$\,K. This effective temperature is significantly cooler than that obtained with the shorter wavelength data. Even though the error of this value is the largest of the three $\teff$ estimates, it is clear that the best fit model to the SWS data underestimates the visible and near-IR data, i.e., the fluxes at wavelengths shorter than about 2\,$\mu$m, as shown in the bottom left corner inset of Fig.~\ref{f:sed_sws}.

\subsubsection{Adopted $\teff$}

\begin{figure*}
\includegraphics[bb=60 370 900 705,width=18cm]{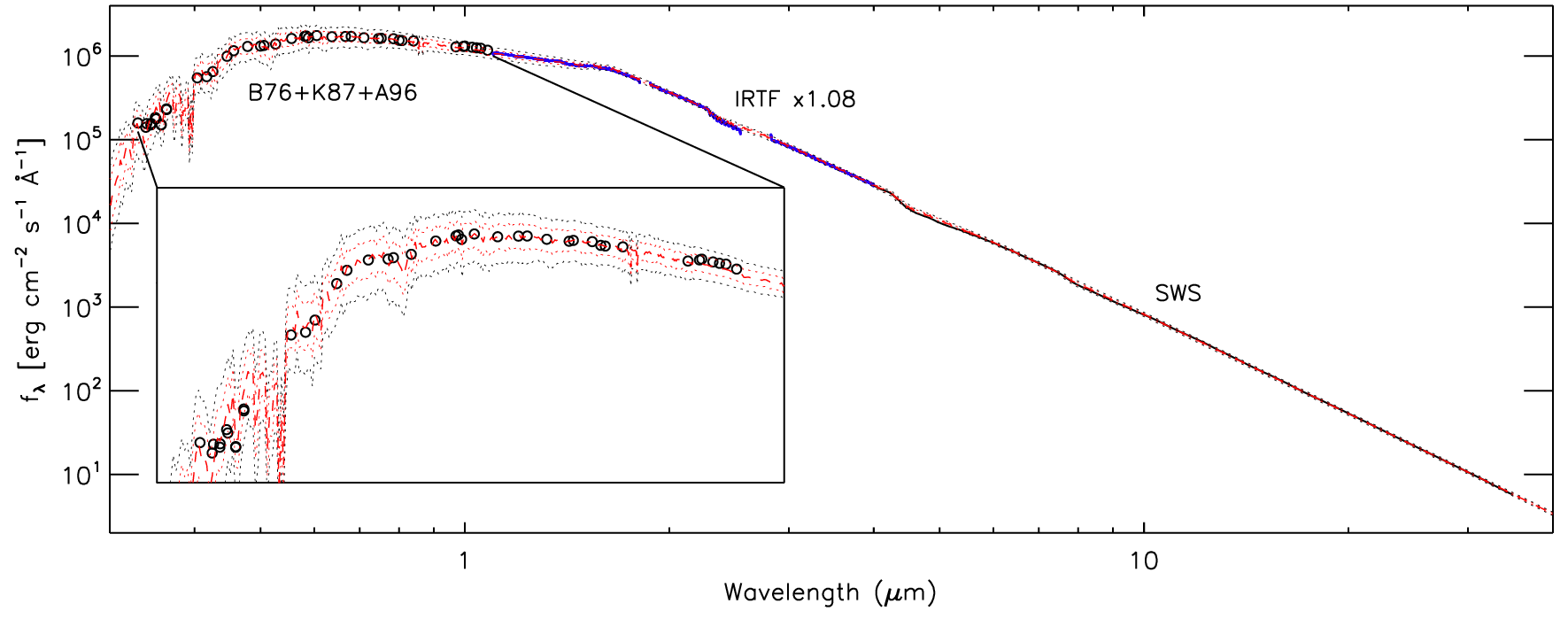}
\caption{Near-UV to mid-infrared spectral energy distribution of Arcturus. Observed data are from \cite{kiehling87}, \cite{breger76}, and \cite{alekseeva96} for the visible (average values shown), IRTF (solid line from 1.1 to 4\,$\mu$m), and SWS (solid line from 4\,$\mu$m). The best fit model is shown with the dotted line. Model fluxes corresponding to $\teff\pm100$\,K and $\teff\pm200$\,K are also shown with dotted lines. The bottom left corner inset shows an expanded view of the 0.3 to 1.2\,$\mu$m region.}
\label{f:sed_all}
\end{figure*}

The effective temperature of Arcturus, as suggested by model fits to observed SEDs in three different spectral regions, is given in Table~\ref{t:teff}. The weighted mean and standard deviation from these three results is: $\teff=4286\pm30$\,K. In Fig.~\ref{f:sed_all} we show the spectral energy distribution of Arcturus from near-UV to mid-infrared wavelengths. The observed data are from different sources. In the visible, we computed an average of all observations available, interpolating the data sets to the wavelengths of the \cite{breger76} data, excluding the \cite{kiehling87} fluxes for wavelength regions affected by telluric absorption. A Kurucz' model flux distribution of $\teff=4286$\,K is also shown in Fig.~\ref{f:sed_all}, along with similar models for $\teff\pm100$\,K and $\teff\pm200$\,K. Clearly, the sensitivity to $\teff$ is more important at shorter wavelengths and this is why the model fits in the near- to mid-infrared regions have a larger uncertainty and therefore much lower weight in the calculation of the final $\teff$. This figure also suggests that our error estimate is realistic. The observed fluxes are fully contained within the $\pm100$\,K model flux limits while inspection of the $\pm50$\,K limits shows that most of the observed visible data are contained within those limits, as shown in the bottom left corner inset of Fig.~\ref{f:sed_all}. Direct integration of the composite SED shown in Fig.~\ref{f:sed_all} results in a bolometric flux (on Arcturus' surface) $f_\mathrm{bol}=\int f_\lambda d\lambda=1.923\times10^{10}$\,erg\,cm$^{-2}$\,s$^{-1}$, which corresponds to an effective temperature $\teff=(f_\mathrm{bol}/\sigma)^{1/4}=4291$\,K ($\sigma$ is the Stefan-Boltzmann constant), which is in excellent agreement with our adopted $\teff$ from the model fits.

\begin{table}
\caption{Effective Temperature of Arcturus from SED Fits}
\label{t:teff}
\centering
\begin{tabular}{ccc}\hline\hline
Spectral & $\teff$ & error \\ 
region & (K) & (K) \\ \hline
Visible & 4288 &  17 \\
IRTF    & 4347 &  69 \\
SWS     & 4152 &  84 \\ \hline
Adopted & 4286 &  30 \\ \hline
\end{tabular}
\end{table}

As explained in Sect.~\ref{s:theta}, our adopted angular diameter of Arcturus could be affected by a systematic error of up to 0.5\,mas, which will have an impact on the calculations presented in this Section. We repeated the SED fits using a larger angular diameter (21.56\,mas) and found $\teff\simeq4260$\,K, i.e., an effective temperature cooler by about 30\,K.\footnote{Note that if $\teff$ were derived directly from the bolometric flux and angular diameter, the change in $\teff$ would be larger, about 50\,K. The value derived here ($\simeq30$\,K) corresponds to the SED fits, which depend not only on the scale factor $s$ but also on the shape of the flux distributions, particularly in the visible, the spectral region that has largest weight when computing the final $\teff$.} We also found slightly larger minimum reduced $\chi^2$ values in all fits. Considering that a 0.5\,mas correction is probably an extreme case, we conclude that the impact of systematic errors on the angular diameter is not crucial for the rest of our analysis.

\subsection{Parallax}

A measurement of the trigonometric parallax of Arcturus is available in the Hipparcos catalog. The new reduction of Hipparcos data \citep{vanleeuwen07} did not change significantly the mean value of the parallax (it went from 88.85 to 88.83\,mas; a distance of 11.3\,pc) but the error bar decreased by almost 30\,\%. The Hipparcos parallaxes are in general much more reliable and precise than ground-based measurements. However, Arcturus is one of the few very bright stars in the catalog and thus the results might suffer from systematic errors. To ensure that the Hipparcos parallax of Arcturus is reliable, we checked ground-based measurements.

Table~\ref{t:parallax} lists independent measurements of the trigonometric parallax of Arcturus. The first three entries are ground-based. Moreover, the value given by \cite{vanaltena95} represents the average of 6 previously published measurements (not including \citealt{harrington93}), which were all in good agreement. Although the average of ground-based measurements appears to be slightly lower ($\simeq88.1$\,mas) compared to the Hipparcos value (88.83\,mas), all measurements are consistent with each other within the estimated 1\,$\sigma$ errors. This suggests that there are no real problems with the Hipparcos parallaxes of very bright stars. Thus, we adopted the weighted average of the parallaxes listed in Table~\ref{t:parallax} as the parallax of Arcturus: $\pi=88.65\pm0.40$\,mas.

\begin{table}
\centering
\caption{Trigonometric Parallax of Arcturus}
\label{t:parallax}
\begin{tabular}{ccl} \hline\hline
$\pi$   & error   & Reference \\
(mas)   & (mas)   & \\ \hline
88.5    & 2.1     & \cite{harrington93} \\
88.4    & 1.8     & \cite{vanaltena95} \\
87.53   & 1.48    & \cite{gatewood08} \\ 
88.83   & 0.54    & \cite{vanleeuwen07} \\ \hline
88.65   & 0.40    & Weighted mean (value adopted) \\ \hline
\end{tabular}
\end{table}

\subsection{Surface Gravity, Age, and Mass} \label{s:logg}

Given that we know the effective temperature of Arcturus with high precision (0.7\,\%) and that its parallax is accurately known (0.5\,\%), the most reliable way to determine the star's surface gravity ($\logg$) consists on placing the star on the HR diagram and comparing its location with theoretical predictions based on stellar evolution calculations. This approach also allows us to estimate the star's mass ($M$) and age ($\tau$).

\begin{figure}
\includegraphics[bb=80 370 450 860,width=8.9cm]{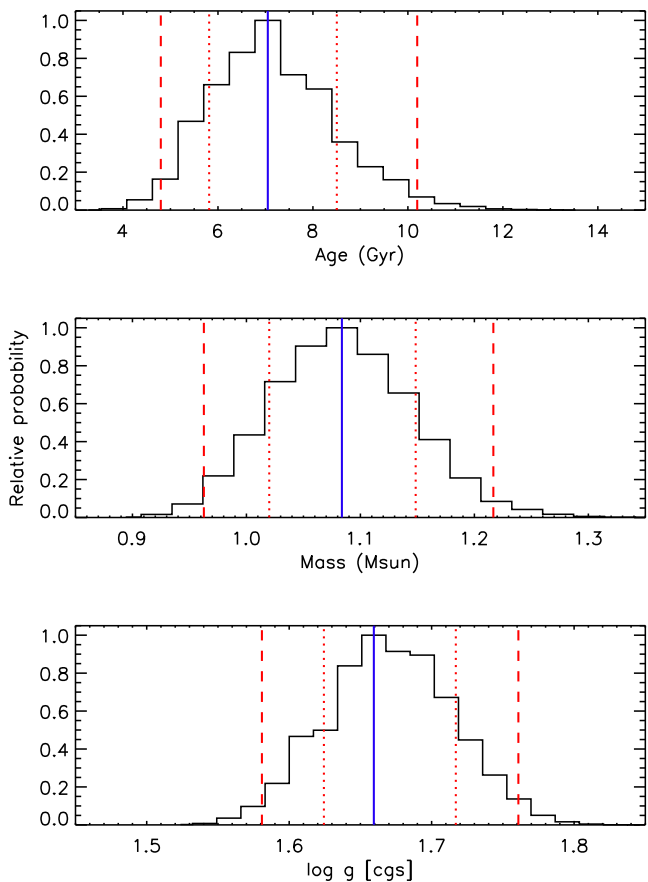}
\caption{Age, mass, and $\logg$ probability distribution functions from our isochrone fit. The vertical solid line is located at the maximum of the PDFs while the location of the Gaussian-like 1\,$\sigma$ and 2\,$\sigma$ lower and upper limits are shown with dotted and dashed lines, respectively.}
\label{f:pdfs}
\end{figure}

We use a classical isochrone fitting technique to determine Arcturus' age, mass, and $\logg$. Details of our particular implementation will be given in a forthcoming paper \citep[][but see also \citealt{reddy03} and \citealt{allende04:s4n}, who use essentially the same approach]{ramirez11:thin-thick}. Briefly, we use a fine grid of Yonsei-Yale isochrones \citep{yi01,kim02}, which consider $\alpha$-element enhancement at sub-solar metallicities, increasing linearly from $[\alpha/\mathrm{Fe}]=0.0$ at $\feh=0.0$ to $[\alpha/\mathrm{Fe}]=+0.3$ at $\feh=-1.0$. We calculate the absolute magnitude of Arcturus from its observed apparent magnitude and trigonometric parallax. The apparent magnitude adopted ($V=-0.051\pm0.013$) is the one listed in the General Catalogue of Photometric Data by \citet[][GCPD]{mermilliod97}, a value that corresponds to a weighted average of more than 89 measurements from 14 independent sources. We checked this value by direct integration of the observed spectral energy distributions used in Sect.~\ref{s:sed_visible}, convolved with the $V$ filter response function by \cite{cohen03}, which ensures a result on the standard Landolt scale. We obtained $V=-0.049\pm0.011$, i.e., fully consistent with the GCPD. For the calculation of the absolute magnitude, apparent magnitude and parallax errors were propagated linearly. We obtain $M_V=-0.313\pm0.016$. We calculate a probability distribution function (PDF) for the isochrone points within 3\,$\sigma$ from the observables $M_V$, $\teff$, and $\feh$. PDFs are computed for the age, mass, and surface gravity, as shown in Fig.~\ref{f:pdfs}, and the values at their maximum are adopted as the stellar parameters. The PDFs are generally asymmetric. Therefore, lower and upper 1\,$\sigma$ and 2\,$\sigma$ Gaussian-like error bars can be determined from the shape of the PDFs. We adopt the 1\,$\sigma$ values as our limits for the error bar estimates. Given the high precision of our derived atmospheric parameters, these PDFs have well defined peaks. The parameters derived from these distributions are: $\tau=7.1^{+1.5}_{-1.2}$\,Gyr, $M=1.08\pm0.06\,M_\odot$, and $\logg=1.66^{+0.6}_{-0.4}$. Hereafter we adopt $\logg=1.66\pm0.05$. We note that these values agree fairly well with some of the early spectroscopic analysis in the literature (e.g., \citealt{ayres77} and references therein).

A number of recent studies have pointed out that the determination of ages and other stellar parameters from isochrones can be severely affected by a number of statistical biases \citep{nordstrom04,pont04,jorgensen05,dasilva06} which can be addressed, for example, using Bayesian methods. We did not take this approach but \cite{dasilva06} provide an online tool to derive the stellar age, mass, and surface gravity using a Bayesian approach from any input observational data.\footnote{http://stev.oapd.inaf.it/cgi-bin/param} We used this tool and found, for our derived parameters of Arcturus: $\tau=8.4\pm2.3$\,Gyr, $M=1.00\pm0.09\,M_\odot$, and $\logg=1.67\pm0.06$. These values are all consistent with our estimates within the quoted 1\,$\sigma$ errors. Note that in addition to improving the method of stellar parameter determination from isochrones, \cite{dasilva06} used a different set of isochrones (those from the Padova group; e.g., \citealt{bertelli94,girardi00}), which could be another source of systematic uncertainty. The excellent agreement between the $\logg$ values that we derive and those obtained using \cite{dasilva06} implementation suggests that the impact of the Bayesian approach and the choice of isochrones is relatively small regarding the determination of surface gravity (and probably mass and age) of stars like Arcturus.

Using our derived surface gravity and mass from isochrones we can calculate the radius of Arcturus: $R=25.2\pm0.4\,R_\odot$. Alternatively, one could use the direct and independent measurements of angular diameter and parallax to calculate this radius: $R=25.5\pm0.2\,R_\odot$. The good agreement between these two estimates of Arcturus' radius gives us confidence that our derived $\logg$ value is not only precise but also accurate. Averaging the two estimates of Arcturus' radius we obtain: $R=25.4\pm0.2\,R_\odot$. If we use an angular diameter larger by 0.5\,mas (Sect.~\ref{s:theta}), we obtain $R=25.6\pm0.4\,R_\odot$ and $R=26.1\pm0.2\,R_\odot$ from the two methods described above. The mass and $\logg$ value were re-computed consistently owing to the lower $\teff$ that corresponds to the larger angular diameter. Thus, in this case we obtain slightly larger radii, which are, however, not as consistent with each other as before.

\subsection{Iron Abundance} \label{s:feh}

The very high signal-to-noise, high resolution spectrum of Arcturus by \cite{hinkle05} was inspected alongside the even higher signal-to-noise, higher resolution solar spectrum by \cite{kurucz84} in order to find spectral lines due to iron (both neutral and singly ionized) which are unblended in both spectra and have a well defined local continuum. Furthermore, we selected only lines that have reliable $\log gf$ values measured in the laboratory. Our adopted $\log gf$ values have been compiled by \cite{ramirez11:thin-thick}. We refer the reader to this paper for details on the sources of laboratory atomic data. The resulting linelist consists of 37~\fei\ lines and 9~\feii\ lines, which are given in Table~\ref{t:feh}, along with the relevant atomic data and equivalent widths ($EW$s) measured in the solar and Arcturus' spectra. The latter were measured using IRAF's task splot,\footnote{IRAF is distributed by the National Optical Astronomy Observatory, which is operated by the Association of Universities for Research in Astronomy, under cooperative agreement with the National Science Foundation.} fitting Gaussian profiles to most lines but Voigt profiles to features with extended wings. Although both the spectra of the Sun and Arcturus are already continuum normalized, in each of our $EW$ measurements we used a local pseudo-continuum determined by visual inspection of a relatively wide spectral window (typically $\pm5$\,\AA\ around each feature). While for the Sun this makes little difference, it can be important for certain spectral regions in the spectrum of Arcturus due to the presence of a large number of weak molecular features. Using the pseudo-continuum minimizes the impact of these small blending features.

\begin{table}
\centering
\caption{Iron Line List}
\label{t:feh}
\begin{tabular}{ccccc}\hline\hline
Wavelength & EP & $\log gf$ & $EW$ Sun & $EW$ Arcturus \\ 
(\AA) & (eV) & & (m\AA) & (m\AA) \\ 
\\
\hline\multicolumn{5}{c}{\fei} \\ \hline
5295.3101 & 4.420 & -1.590 & 29.0 &  47.7 \\
5379.5698 & 3.690 & -1.510 & 62.5 &  99.0 \\
5386.3301 & 4.150 & -1.670 & 32.6 &  56.0 \\
5441.3398 & 4.310 & -1.630 & 32.5 &  55.7 \\
5638.2598 & 4.220 & -0.770 & 80.0 & 104.9 \\
5679.0229 & 4.652 & -0.750 & 59.6 &  72.7 \\
5705.4639 & 4.301 & -1.355 & 38.0 &  62.7 \\
5731.7598 & 4.260 & -1.200 & 57.7 &  83.0 \\
5778.4531 & 2.588 & -3.440 & 22.1 &  74.7 \\
5793.9141 & 4.220 & -1.619 & 34.2 &  56.6 \\
5855.0762 & 4.608 & -1.478 & 22.4 &  37.1 \\
5905.6699 & 4.650 & -0.690 & 58.6 &  74.7 \\
5927.7900 & 4.650 & -0.990 & 42.9 &  56.3 \\
5929.6802 & 4.550 & -1.310 & 40.0 &  57.1 \\
6003.0098 & 3.880 & -1.060 & 84.0 & 112.4 \\
6027.0498 & 4.076 & -1.090 & 64.2 &  93.6 \\
6056.0000 & 4.730 & -0.400 & 72.6 &  85.2 \\
6079.0098 & 4.650 & -1.020 & 45.6 &  60.2 \\
6093.6440 & 4.607 & -1.300 & 30.9 &  45.9 \\
6096.6650 & 3.984 & -1.810 & 37.6 &  64.7 \\
6151.6182 & 2.176 & -3.282 & 49.8 & 119.7 \\
6165.3599 & 4.143 & -1.460 & 44.8 &  70.6 \\
6187.9902 & 3.940 & -1.620 & 47.6 &  78.4 \\
6240.6460 & 2.223 & -3.287 & 48.2 & 118.7 \\
6270.2251 & 2.858 & -2.540 & 52.4 & 107.9 \\
6703.5669 & 2.759 & -3.023 & 36.8 &  92.7 \\
6705.1021 & 4.607 & -0.980 & 46.4 &  64.3 \\
6713.7451 & 4.795 & -1.400 & 21.2 &  30.6 \\
6726.6670 & 4.607 & -1.030 & 46.9 &  62.4 \\
6793.2588 & 4.076 & -2.326 & 12.8 &  29.4 \\
6810.2632 & 4.607 & -0.986 & 50.0 &  66.4 \\
6828.5898 & 4.640 & -0.820 & 55.9 &  72.2 \\
6842.6899 & 4.640 & -1.220 & 39.1 &  55.0 \\
6843.6602 & 4.550 & -0.830 & 60.9 &  80.0 \\
6999.8799 & 4.100 & -1.460 & 53.9 &  78.8 \\
7022.9502 & 4.190 & -1.150 & 64.5 &  89.9 \\
7132.9902 & 4.080 & -1.650 & 43.1 &  68.6 \\
\hline\multicolumn{5}{c}{\feii} \\ \hline
4576.3330 & 2.844 & -2.950 & 64.6 &  76.7 \\
4620.5132 & 2.828 & -3.210 & 50.4 &  60.1 \\
5234.6240 & 3.221 & -2.180 & 82.9 &  88.7 \\
5264.8042 & 3.230 & -3.130 & 46.1 &  47.9 \\
5414.0718 & 3.221 & -3.580 & 27.3 &  31.0 \\
5425.2568 & 3.200 & -3.220 & 41.9 &  45.4 \\
6369.4619 & 2.891 & -4.110 & 19.2 &  23.1 \\
6432.6758 & 2.891 & -3.570 & 41.3 &  47.0 \\
6516.0771 & 2.891 & -3.310 & 54.7 &  59.0 \\
\hline
\end{tabular}
\end{table}

The latest versions of Kurucz' no-overshoot model atmospheres and the spectrum synthesis code MOOG \citep[e.g.,][]{sneden73} were used to compute elemental abundances. For Arcturus, we adopted a model atmosphere with $\alpha$-element enhancement ($[\alpha/\mathrm{Fe}]=+0.4$). van der Waals damping constants were adopted from the works by \cite{barklem00} and \cite{barklem05}.

Abundances were measured differentially, on a line-by-line basis, with respect to the Sun. In every abundance calculation described below, a microturbulent velocity ($v_t$) was derived so that the abundances do not correlate with the reduced equivalent widths ($\mathrm{REW}=\log EW/\lambda$) of the \fei\ lines. The microturbulent velocity is thus different for each case tested but this has a minor impact on the results and the discussion below.

\begin{figure}
\includegraphics[bb=70 370 450 685,width=8.6cm]{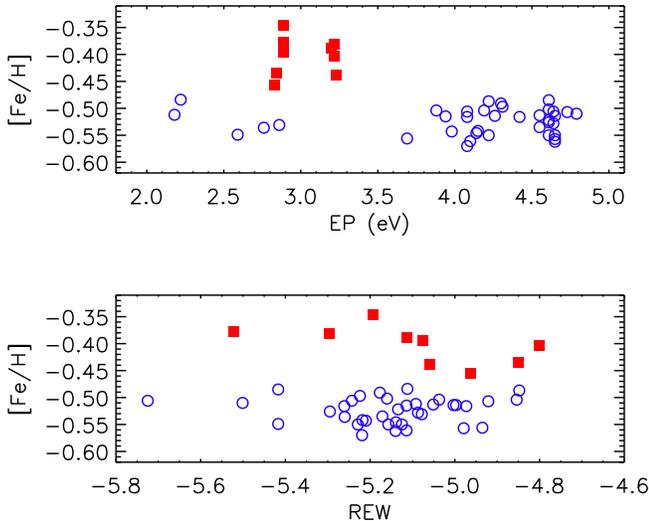}
\caption{Iron abundance of Arcturus, differential on a line-by-line basis with respect to the Sun, as a function of excitation potential and reduced equivalent width of the lines. Neutral iron lines (\fei) are represented with open circles while singly ionized iron lines (\feii) are shown with filled squares.}
\label{f:feabund_diff}
\end{figure}

Using our derived values of $\teff=4286$\,K and $\logg=1.66$, we obtain a mean $\feh=-0.52\pm0.02$ from the \fei\ lines and $\feh=-0.40\pm0.03$ from the \feii\ lines (see Fig.~\ref{f:feabund_diff}; the derived microturbulent velocity is $v_t=1.74$~km~s$^{-1}$). The error bars here correspond only to the line-to-line scatter and do not include the uncertainty introduced by errors in $\teff$ and $\logg$. The mean iron abundances that we obtain from \fei\ and \feii\ lines separately are thus inconsistent by 0.12\,dex. On the other hand, the iron abundances inferred from the \fei\ lines do not correlate significantly with excitation potential.

\begin{figure}
\includegraphics[bb=70 370 390 685,width=8.8cm]{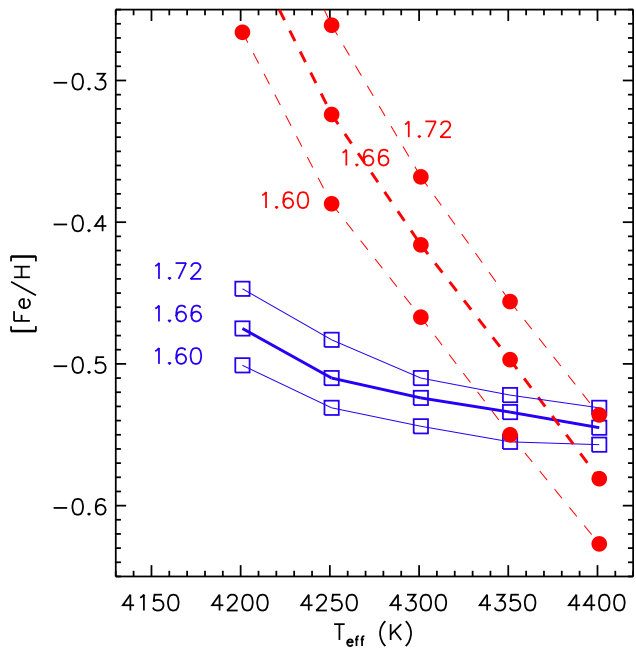}
\caption{Iron abundance derived from \fei\ (open squares connected by solid lines) and \feii\ (filled circles connected by dashed lines) spectral lines as a function of input $\teff$. Three possible choices of $\logg=1.60,1.66,1.72$ are tested. Thick lines correspond to our preferred value of $\logg$.}
\label{f:speq1}
\end{figure}

In Fig.~\ref{f:speq1} we show the iron abundance inferred from \fei\ and \feii\ lines as a function of $\teff$ for various values of $\logg$. Clearly, the \feii\ lines are more sensitive to both parameters. It is not possible to reconcile the \fei\ and \feii\ abundances within our 1D-LTE approach for
the effective temperature derived in Section~\ref{s:teff}. Further work on the effects of non-LTE and possibly also surface inhomogeneities is needed to tackle this problem, but this is beyond the scope of this paper. Ionization balance would be achieved if the effective temperature of Arcturus were about 4380\,K, which is inconsistent with our preferred value for this parameter (our derived $\teff$ would have to be off by more than 3\,$\sigma$).

\begin{figure}
\includegraphics[bb=70 370 390 685,width=8.7cm]{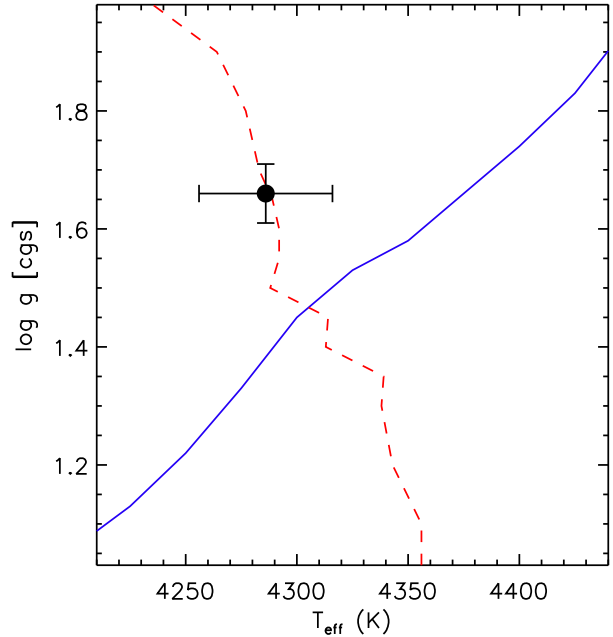}
\caption{The location of $\teff,\logg$ pairs for which ionization balance is satisfied is shown with the solid line. Similarly, the location of $\teff,\logg$ pairs for which excitation balance is satisfied is shown with the dashed line. The filled circle with error bars corresponds to our derived values of $\teff$ and $\logg$ for Arcturus, obtained independently from the iron line analysis.}
\label{f:speq2}
\end{figure}

The loci of stellar parameters $\teff$ and $\logg$ for which the conditions of ionization and excitation balance are satisfied are shown in Fig.~\ref{f:speq2}. This figure shows that the iron abundance analysis with our preferred $\teff$ and $\logg$ values satisfies excitation balance but not ionization balance. Thus, at least within our 1D-LTE approach, the analysis of \fei\ lines seems more reliable than that for the \feii\ lines. Note that at the temperatures found in the line-forming layers of photospheres of stars like Arcturus, iron is found mostly in its neutral stage, thus making \fei\ lines more robust against departures from LTE and errors in the atmospheric parameters. On the other hand, \feii\ dominates in the layers where the continuum is formed. We adopt the iron abundance from \fei\ lines as the iron abundance of Arcturus because it is internally more robust and consistent with the rest of our 1D-LTE analysis. The discrepancy regarding the \feii\ line analysis cannot be solved within the 1D-LTE approach; it must be addressed in future work but for now it should be accepted as one of the limitations of the standard spectroscopic approach for the determination of iron abundances.\footnote{The possible systematic error of 0.5\,mas in the angular diameter (Sect.~\ref{s:theta}) would not solve these problems. The decrease of about 30\,K in $\teff$ worsens the ionization imbalance and it is not compensated enough by the small lowering of the $\logg$ value, which is about 0.04\,dex.}

Hereafter, the value of iron abundance that we adopt is $\feh=-0.52\pm0.04$. In this case we have also included the error introduced by the uncertain $\teff$ and $\logg$ values, in addition to the line-by-line scatter.

The results described above do not change if we use absolute abundances instead of differential ones. However, the internal errors are reduced significantly if we use differential analysis. This suggests that the only advantage of using a solar spectrum as reference for differential analysis of Arcturus (and probably other red giant stars) is to minimize the errors due to uncertain $\log gf$ values. Systematic errors due to simplifications in the model atmosphere computations and/or spectral line synthesis are not removed with a solar differential analysis.

\subsection{Kurucz vs.\ MARCS Model Atmospheres}

For consistency, in this paper we use only plane-parallel Kurucz model atmospheres. However, we repeated the entire procedure described in this section to derive the atmospheric parameters using the latest MARCS model atmosphere grid \citep{gustafsson08} with standard chemical composition, which means that $\alpha$-element enhancement is considered for stars with $\feh<0$.\footnote{The MARCS grid of atmospheric models used in this work is available online at http://marcs.astro.uu.se} Plane-parallel MARCS models are only available for $\logg>3.0$ so for this analysis of Arcturus we used spherically symmetric MARCS models. We find: $\teff=4282\pm36$\,K, $\logg=1.66\pm0.05$, and $\feh=-0.54\pm0.05$, i.e., fully consistent with the results obtained with Kurucz model atmospheres. The iron abundance analysis was also made differentially on a line-by-line basis with respect to the Sun.

The discussion regarding ionization balance using Kurucz models, given in the previous section, does not change qualitatively for MARCS models. The mean \feii\ abundance in this case is also about 0.1\,dex higher than that from the \fei\ lines. The only important difference with respect to the Kurucz model atmosphere analysis is that the excitation balance is no longer satisfied (hence the slightly large error for $\feh$). For a $\logg=1.66$, the effective temperature has to increase to about 4370\,K if MARCS models are used. Kurucz models, on the other hand, are consistent with excitation balance if $\teff\simeq4290$\,K, as suggested by the spectral energy distribution. Despite this discrepancy, the atmospheric parameters derived are nearly independent on which set of the most commonly used model atmosphere grids, namely Kurucz and MARCS, are used. Note that the excitation balance is heavily sensitive to the 5 \fei\ lines with $\mathrm{EP}<3$\,eV whereas many more high excitation potential lines are available. Small errors in the measured equivalent widths could affect somewhat these results. Thus, we must not imply a superiority of one set of models over the other based on our \fei\ line analysis.

\section{Elemental Abundances}

We employed a curve-of-growth (COG) approach, using MOOG, to measure elemental abundances of 16 elements other than iron. Equivalent widths were measured carefully with IRAF's task splot, de-blending lines when necessary. For each element other than K, more than one line was available. The weighted average abundance was finally adopted, with the weights being determined from the actual COGs (i.e., from the slope of the abundance versus reduced equivalent width relation, assuming a typical error of 1\,\% for the latter\footnote{Formal $EW$ errors can be computed from the properties of the observed spectra, as in \cite{cayrel88}. We find errors of order 0.1\,m\AA, which is about 0.2\,\% for an $EW=50$\,m\AA\ line. We prefer to adopt a larger error because of the uncertainties introduced by the continuum placement, which roughly scale with line strength. At this point the error bars are used only to weight the abundances inferred from different lines of the same element so our results will not be significantly affected if these errors are under- or over-estimated, as
long as their relative values are correct.}). Thus, strong and possibly saturated lines, in particular those in the spectrum of Arcturus, are given lower weight (in this way we also minimize the impact of uncertain broadening parameters; see below). Similar to the case of iron, we calculated abundances on a line-by-line basis and differentially with respect to the Sun before averaging, in order to reduce the impact of errors in the atomic data (with a few exceptions, as explained below). This implies that not all available lines were used but only those for which reliable equivalent widths could be measured in the spectra of both Arcturus and the Sun. Our measured equivalent widths, atomic line data, and absolute abundances are given in Table~\ref{t:linelist}. The abundance errors in this table correspond only to those from the COG analysis and do not include the uncertainties from the stellar parameters.

\begin{table*}
\centering
\caption{Line List for Elements other than Iron}
\begin{tabular}{l|ccc|ccc|ccc|cc}\hline\hline
& & & & \multicolumn{3}{c}{Sun} & \multicolumn{3}{c}{Arcturus} & & \\
Species & Wavelength & EP & $\log gf$ & EW & $A_\mathrm{X}$ & error & EW & $A_\mathrm{X}$ & error & [X/H] & error \\ 
& (\AA) & (eV) & & (m\AA) & & & (m\AA) & & & & \\ \hline
C\,\textsc{i} & 5380.34 & 7.68 &  -1.62 &  21.3 & 8.42 & 0.01 &   9.4 & 8.35 & 0.01 & -0.07 & 0.02 \\
C\,\textsc{i} & 8335.15 & 7.68 &  -0.44 &  89.5 & 8.37 & 0.02 &  29.0 & 8.32 & 0.02 & -0.05 & 0.03 \\
C\,\textsc{i} & 9078.28 & 7.48 &  -0.57 & 106.3 & 8.47 & 0.02 &  33.1 & 8.30 & 0.02 & -0.17 & 0.03 \\
C\,\textsc{i} & 9111.80 & 7.49 &  -0.30 & 128.1 & 8.40 & 0.02 &  45.0 & 8.32 & 0.02 & -0.08 & 0.03 \\

O\,\textsc{i} & 5577.34 & 1.97 &  -8.24 & \nodata      & 8.69 & \nodata     &  10.5 & 8.68 & 0.01 & -0.01 & 0.01 \\
O\,\textsc{i} & 6300.30 & 0.00 &  -9.72 & \nodata      & 8.69 & \nodata     &  68.1 & 8.66 & 0.02 & -0.03 & 0.02 \\
O\,\textsc{i} & 6363.78 & 0.02 & -10.19 & \nodata      & 8.69 &  \nodata    &  31.1 & 8.66 & 0.01 & -0.03 & 0.01 \\

Na\,\textsc{i} & 4751.82 & 2.10 &  -2.08 &  11.5 & 6.20 & 0.01 &  29.0 & 5.77 & 0.01 & -0.43 & 0.01 \\
Na\,\textsc{i} & 5148.84 & 2.10 &  -2.04 &  12.9 & 6.21 & 0.01 &  35.5 & 5.83 & 0.01 & -0.38 & 0.02 \\
Na\,\textsc{i} & 6154.23 & 2.10 &  -1.55 &  37.0 & 6.26 & 0.01 &  73.3 & 5.84 & 0.02 & -0.42 & 0.03 \\
Na\,\textsc{i} & 6160.75 & 2.10 &  -1.25 &  55.8 & 6.25 & 0.02 &  92.9 & 5.82 & 0.03 & -0.43 & 0.03 \\
 
\ldots & \ldots & \ldots & \ldots & \ldots & \ldots & \ldots & \ldots & \ldots & \ldots & \ldots & \ldots \\ \hline
\end{tabular}
\label{t:linelist}
\end{table*}

Inspection of Table~\ref{t:linelist} reveals that the spectral lines in the spectrum of Arcturus tend to be stronger than those in the solar spectrum. This is also the case of the iron lines employed. Pressure broadening could affect significantly the $EW$s of strong lines while being less important for weak features. Thus, it is necessary to investigate possible errors introduced by this ingredient on the line formation calculations. We re-computed the abundances using the classical formula by Uns\"old to calculate the van der Waals damping constants instead of using those by \cite{barklem00}. On a line-by-line basis, the differential iron abundances increased between 0.00 and 0.04\,dex, depending on the line, while the average $\feh$ increased by 0.02\,dex. For the abundance ratios, [X/Fe], we obtained shifts of only $\pm0.02$\,dex. It has been shown that the Uns\"old approximation severely underestimates the damping constants while the calculations by \cite{barklem00} improve abundance determinations. Thus, the differences quoted here, albeit small, should still not be considered as potential systematic errors. They only represent the worst case scenario of errors introduced by uncertain damping constants. Other pressure broadening prescriptions will result in abundance ratios much more similar to those obtained using \cite{barklem00} constants.

Only atomic lines have been used for these calculations even though the spectra of cool giant stars like Arcturus are rich in molecular features. We prefer to avoid the latter in our work because of the potentially severe errors that could be introduced by surface inhomogeneities which are not taken into account in the modeling of the star's atmosphere \citep[e.g.,][]{asplund05:review}. It has been shown that these so-called 3D effects are very important for molecular features in cool giants, an effect that is further enhanced by a low metallicity \citep[e.g.,][]{collet07,collet09}. While these studies are mostly theoretical, some fundamental predictions of these 3D models have been tested against high quality observations, showing in general good agreement and therefore providing support to the findings of large 3D abundance corrections \citep[e.g.,][]{ramirez10:granulation}.

Hyperfine splitting was taken into account for the analysis of Mn and Co lines. The equivalent widths of these features could be measured with high precision, so line profile fitting was not necessary. Instead, we used the ``blends'' driver in MOOG to take the effect into account. Hyperfine structure constants were adopted from Kurucz\footnote{http://kurucz.harvard.edu} but with the transition probabilities scaled to the total $\log gf$ values given in \cite{asplund09:review}.

Nearly all spectral lines used in this work come from the line selection by \cite{asplund09:review}, therefore ensuring that our solar reference analysis is as accurate as possible. Indeed, the average difference between our solar abundances and those by \citeauthor{asplund09:review}, excluding oxygen (see below), is $\Delta(A_\mathrm{X})=-0.014\pm0.082$. The largest difference is seen for K, for which we find $A_\mathrm{K}=5.31$ whereas \citeauthor{asplund09:review} derive $A_\mathrm{K}=5.03$. This element has only one very strong line available for analysis so its abundance as reported in this work should not be considered very reliable. In fact, \cite{zhang06} report a non-LTE correction of $-0.29$ for this line, which would bring our K abundance into excellent agreement with the value derived by \citeauthor{asplund09:review} Excluding K, the average difference with \citeauthor{asplund09:review} solar abundances is $\Delta(A_\mathrm{X})=-0.033\pm0.033$; i.e., our solar abundances are marginally lower. We note that, in general, the differences between our solar abundances and those by \citeauthor{asplund09:review} can be reconciled if we take non-LTE effects into account, as they did.\footnote{They also consider 3D effects but the 3D--1D differences are modest for atomic lines.} The latter are dominated by over-ionization if weak and moderately strong lines are used, resulting in lower abundances if derived from neutral species, as is our case (with few exceptions).

Our adopted atomic data are also from the compilation by \cite{asplund09:review}. Their line selection is very strict, therefore leaving us with some elements with very few or no measurable lines available in both the solar and Arcturus' spectra. This was the case of C and Al. For these elements we complemented the linelist with atomic data from the NIST\footnote{http://physics.nist.gov/PhysRefData/ASD/lines\_form.html} database (for the 8335, 9078, and 9112\,\AA\ C\,\textsc{i} lines) and from \cite{melendez09:twins} for the 7836\,\AA\ Al\,\textsc{i} line. The oxygen lines used here are not easy to measure in the solar spectrum because they are very weak and heavily blended; detailed line synthesis must be employed for those features. For these lines, we adopted the average solar oxygen abundance given by \citeauthor{asplund09:review} ($A_\mathrm{O}=8.69$).

The weighted average abundances relative to the solar abundances we derive for Arcturus are listed in Table~\ref{t:abund_final}, where we also summarize our error analysis. The contribution to the error by the line-to-line scatter is given by $\sigma_\mathrm{line}$, and corresponds to the standard error of the line-by-line abundance dispersion for each element. The errors introduced by our formal uncertainties in the atmospheric parameters are given by $\sigma_t$ (corresponding to the $\teff$ uncertainty), $\sigma_g$ (for $\logg$), and $\sigma_m$ (for $\feh$). They were computed using the abundances derived from model atmospheres with slightly modified stellar parameters compared to those obtained with our preferred values. The four error contributions were then added in quadrature to obtain the final error estimates, which are given in the last column of Table~\ref{t:abund_final}. This is equivalent to assuming that the covariances between the uncertainties in the atmospheric parameters are small, which is approximately true given the procedures used to constrain them.

\begin{table}
\footnotesize
\centering
\caption{Error Analysis and Mean Elemental Abundances of Arcturus}
\begin{tabular}{lcccccr}\hline\hline
Species & [X/H] & $\sigma_\mathrm{line}$ & $\sigma_t$ & $\sigma_g$ & $\sigma_m$ & [X/Fe] \\ \hline
C\,\textsc{i} & -0.09 & 0.03 & 0.050 & 0.035 & 0.005 & $ 0.43\pm0.07$ \\
O\,\textsc{i} & -0.02 & 0.02 & 0.000 & 0.020 & 0.010 & $ 0.50\pm0.03$ \\
Na\,\textsc{i} & -0.41 & 0.02 & 0.025 & 0.000 & 0.000 & $ 0.11\pm0.03$ \\
Mg\,\textsc{i} & -0.15 & 0.03 & 0.005 & 0.010 & 0.005 & $ 0.37\pm0.03$ \\
Al\,\textsc{i} & -0.18 & 0.03 & 0.015 & 0.000 & 0.005 & $ 0.34\pm0.03$ \\
Si\,\textsc{i} & -0.19 & 0.02 & 0.025 & 0.015 & 0.005 & $ 0.33\pm0.04$ \\
K\,\textsc{i} & -0.32 & 0.05 & 0.045 & 0.005 & 0.005 & $ 0.20\pm0.07$ \\
Ca\,\textsc{i} & -0.41 & 0.02 & 0.035 & 0.005 & 0.005 & $ 0.11\pm0.04$ \\
Sc\,\textsc{i} & -0.37 & 0.06 & 0.045 & 0.000 & 0.000 & $ 0.15\pm0.08$ \\
Sc\,\textsc{ii} & -0.29 & 0.03 & 0.005 & 0.025 & 0.015 & $ 0.23\pm0.04$ \\
Ti\,\textsc{i} & -0.25 & 0.02 & 0.050 & 0.000 & 0.000 & $ 0.27\pm0.05$ \\
Ti\,\textsc{ii} & -0.31 & 0.03 & 0.005 & 0.025 & 0.010 & $ 0.21\pm0.04$ \\
V\,\textsc{i} & -0.32 & 0.02 & 0.050 & 0.005 & 0.005 & $ 0.20\pm0.05$ \\
Cr\,\textsc{i} & -0.57 & 0.03 & 0.025 & 0.000 & 0.000 & $-0.05\pm0.04$ \\
Mr\,\textsc{i} & -0.73 & 0.03 & 0.020 & 0.005 & 0.005 & $-0.21\pm0.04$ \\
Co\,\textsc{i} & -0.43 & 0.03 & 0.000 & 0.015 & 0.010 & $ 0.09\pm0.04$ \\
Ni\,\textsc{i} & -0.46 & 0.02 & 0.005 & 0.015 & 0.005 & $ 0.06\pm0.03$ \\
Zn\,\textsc{i} & -0.30 & 0.05 & 0.030 & 0.025 & 0.005 & $ 0.22\pm0.06$ \\
 \hline
\end{tabular}
\label{t:abund_final}
\end{table}

For Sc and Ti, spectral lines from two species, namely neutral and singly ionized, were available for our abundance analysis. In principle in both cases ionization balance is satisfied within the $1\sigma$ uncertainties. The average abundance for the two Ti species leads to $\mathrm{[Ti/Fe]}=0.23\pm0.04$. Nevertheless, it is clear that the line-to-line scatter for the Sc\,\textsc{i} abundance is much larger than that for Sc\,\textsc{ii}. It is known that, for solar-type stars, Sc\,\textsc{i} abundances are severely affected by non-LTE effects \cite[e.g.,][]{zhang08} whereas those obtained from Sc\,\textsc{ii} lines are more reliable. The smaller error of the latter suggests that this could be the case also for giant stars. Therefore, for the Sc abundance we prefer to use that inferred from Sc\,\textsc{ii} lines exclusively.

\begin{figure*}
\centering
\includegraphics[bb=75 360 900 1085,width=18.2cm]{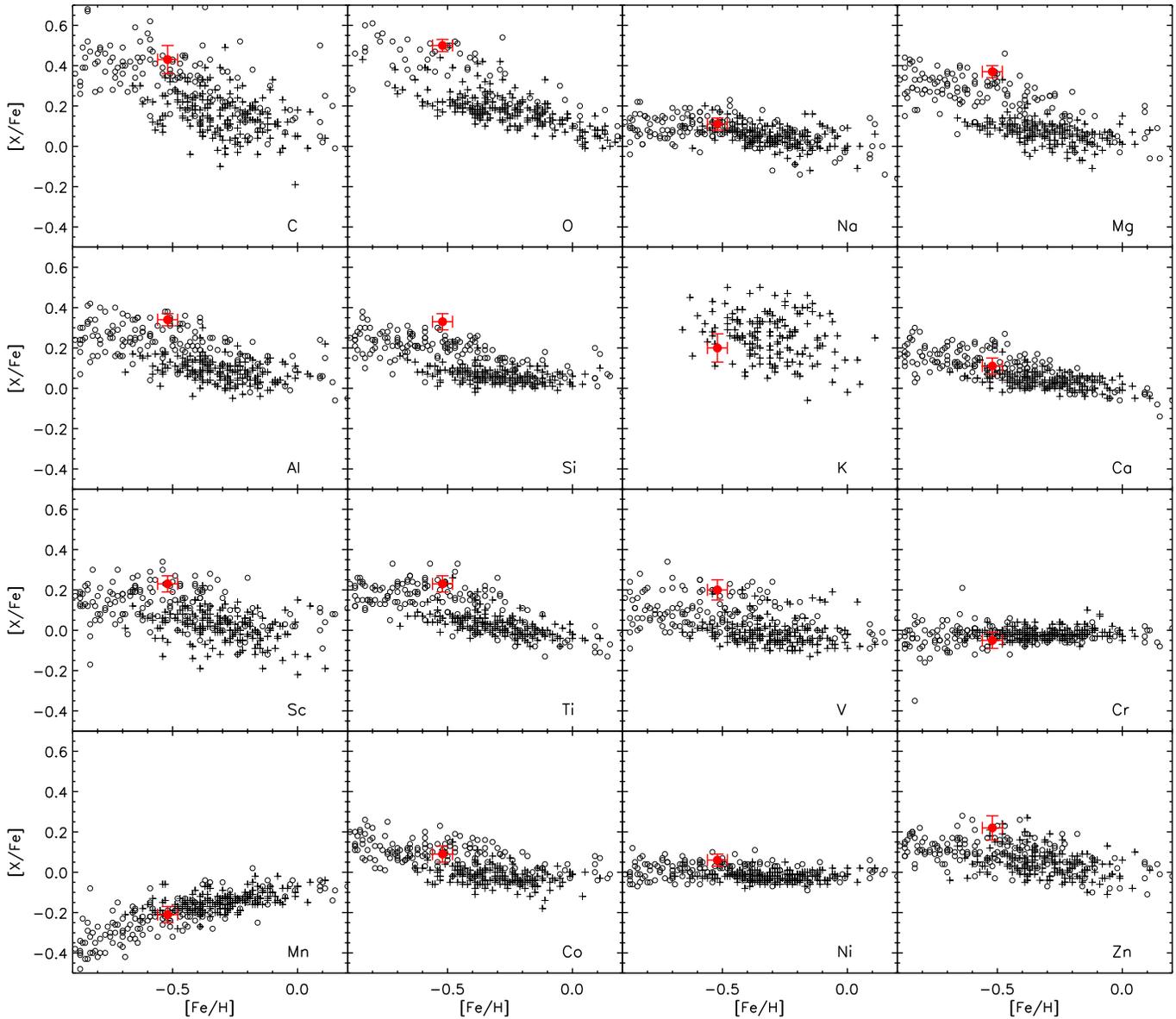}
\caption{Elemental abundance ratios of Arcturus (filled circle with error bars) compared to the Galactic chemical evolution trends by \cite{reddy03} for the thin disk (crosses)  and \cite{reddy06} for the thick disk (open circles), except for the oxygen trend, which is by \cite{ramirez07}. The membership criterion adopted in these studies is purely kinematic.}
\label{f:reddy}
\end{figure*}

Fig.~\ref{f:reddy} shows our abundance results compared to Galactic chemical evolution trends based on the analysis of large samples of nearby dwarf stars from \cite{reddy03,reddy06} and \cite{ramirez07}. These trends separate stars that are members of the so-called Galactic thin disk from those of the thick disk \cite[e.g.,][]{fuhrmann98,bensby03,reddy03,reddy06}. The latter tend to have hotter kinematics compared to the thin disk \cite[e.g.,][]{soubiran93} and based on their observed properties (kinematics, chemical composition, age distribution, etc.), they are thought to be a different stellar population, separate from the thin disk, although its precise origin remains unknown. The (heliocentric) Galactic space velocity components of Arcturus, as derived by \cite{ramirez11:thin-thick}, are $U=25.2\pm1.3$\,\kms, $V=-119.0\pm1.0$\,\kms, $W=-2.7\pm3.5$\,\kms. Using the membership formulation by \cite{ramirez07}, we find that Arcturus has a probability of about 94\,\% of being a thick-disk member.

As shown in Fig.~\ref{f:reddy}, the chemical composition of Arcturus is very typical of that of a thick-disk star in the solar neighborhood. The enhanced abundances we obtain for C, O, Mg, Al, Si, Sc, Ti, and V are also observed in ordinary thick-disk dwarf stars. The Ca abundance appears slightly low compared to the mean thick disk trend but it is still fully consistent with it considering the $1\sigma$ star to star scatter. We note that our K abundance appears somewhat low compared to the mean thin-disk trend (no K abundances were derived by \citealt{reddy06} for thick-disk stars), but we recall that the abundance of this element is based on the analysis of only one not so reliable feature. Relatively small abundance offsets between Arcturus and the mean thick-disk trends in this comparison are not unreasonable given the important differences in stellar parameters between Arcturus and dwarf stars.

\section{Conclusions}

Atmospheric parameters and elemental abundances of Arcturus, a primer for studies of red giant stars, are derived using high quality data and methods that minimize model uncertainties within the 1D-LTE approach. Limitations of these techniques are revealed in the inconsistency between the iron abundance inferred from \fei\ and \feii\ lines and possibly also in the large error of Sc\,\textsc{i} abundances. These cases should be addressed using more sophisticated models, such as those including the impact of surface inhomogeneities and non-LTE.

The elemental abundance pattern of Arcturus is typical of that observed in ordinary nearby thick-disk stars and consistent with its space velocity and relatively old age. \cite{navarro04} suggested an extragalactic origin for the Arcturus group based on its angular momentum. Elemental abundance studies of stars in the Milky Way's satellite galaxies, however, show that extragalactic stars could be disentangled from the Galactic disk and halo stars using their detailed abundance patterns. For example, their [Mg/Fe] and [Ca/Fe] abundance ratios should be significantly lower than those observed in the solar neighborhood \cite[e.g.,][]{venn04,tolstoy09}. The elemental abundances we derive suggest that Arcturus was born in the Milky Way.

\acknowledgments

I.R.'s work was performed under contract with the California Institute of Technology (Caltech) funded by NASA through the Sagan Fellowship Program.


\begin{thebibliography}{81}
\expandafter\ifx\csname natexlab\endcsname\relax\def\natexlab#1{#1}\fi

\bibitem[{{Alekseeva} {et~al.}(1996){Alekseeva}, {Arkharov}, {Galkin},
  {Hagen-Thorn}, {Nikanorova}, {Novikov}, {Novopashenny}, {Pakhomov}, {Ruban},
  \& {Shchegolev}}]{alekseeva96}
{Alekseeva}, G.~A., {et~al.} 1996, Baltic Astronomy, 5, 603

\bibitem[{Allende Prieto} {et~al.}(2002)]{allende02} {Allende Prieto}, C., {Asplund}, M., {Garc{\'{\i}}a L{\'o}pez}, R.~J.,  \& {Lambert}, D.~L. 2002, \apj, 567, 544

\bibitem[{{Allende~Prieto} {et~al.}(2004){Allende~Prieto}, {Barklem},
  {Lambert}, \& {Cunha}}]{allende04:s4n}
{Allende~Prieto}, C., {Barklem}, P.~S., {Lambert}, D.~L., \& {Cunha}, K. 2004,
  \aap, 420, 183

\bibitem[{{Allende Prieto} {et~al.}(2008){Allende Prieto}, {Majewski},
  {Schiavon}, {Cunha}, {Frinchaboy}, {Holtzman}, {Johnston}, {Shetrone},
  {Skrutskie}, {Smith}, \& {Wilson}}]{allende08:apogee}
{Allende Prieto}, C., {et~al.} 2008, Astronomische Nachrichten, 329, 1018

\bibitem[{{Alves-Brito} {et~al.}(2010){Alves-Brito}, {Mel{\'e}ndez}, {Asplund},
  {Ram{\'{\i}}rez}, \& {Yong}}]{alves-brito10}
{Alves-Brito}, A., {Mel{\'e}ndez}, J., {Asplund}, M., {Ram{\'{\i}}rez}, I., \&
  {Yong}, D. 2010, \aap, 513, A35

\bibitem[{{Asplund}(2005)}]{asplund05:review}
{Asplund}, M. 2005, \araa, 43, 481

\bibitem[{{Asplund} {et~al.}(2009){Asplund}, {Grevesse}, {Sauval}, \&
  {Scott}}]{asplund09:review}
{Asplund}, M., {Grevesse}, N., {Sauval}, A.~J., \& {Scott}, P. 2009, \araa, 47,
  481

\bibitem[{Ayres}(1986)]{ayres86} {Ayres}, T.~R. 1986, \apj, 308, 246

\bibitem[{{Ayres}(2010)}]{ayres10}
{Ayres}, T.~R. 2010, \apjs, 187, 149

\bibitem[{{Ayres} \& {Johnson}}(1977)]{ayres77} {Ayres}, T.~R., \& {Johnson}, H.~R. 1977, \apj, 214, 410

\bibitem[{Ayres} \& {Linsky}(1975)]{ayres75} {Ayres}, T.~R., \& {Linsky}, J.~L. 1975, \apj, 200, 660

\bibitem[{{Barklem} \& {Aspelund-Johansson}(2005)}]{barklem05}
{Barklem}, P.~S., \& {Aspelund-Johansson}, J. 2005, \aap, 435, 373

\bibitem[{{Barklem} {et~al.}(2000){Barklem}, {Piskunov}, \&
  {O'Mara}}]{barklem00}
{Barklem}, P.~S., {Piskunov}, N., \& {O'Mara}, B.~J. 2000, \aaps, 142, 467

\bibitem[{{Bensby} {et~al.}}(2003)]{bensby03} {Bensby}, T., {Feltzing}, S., \& {Lundstr{\"o}m}, I. 2003, \aap, 410, 527

\bibitem[{{Bertelli} {et~al.}(1994){Bertelli}, {Bressan}, {Chiosi}, {Fagotto},
  \& {Nasi}}]{bertelli94}
{Bertelli}, G., {Bressan}, A., {Chiosi}, C., {Fagotto}, F., \& {Nasi}, E. 1994,
  \aaps, 106, 275

\bibitem[{{Breger}(1976)}]{breger76}
{Breger}, M. 1976, \apjs, 32, 7

\bibitem[{{Castelli} \& {Kurucz}(2003)}]{castelli03}
{Castelli}, F., \& {Kurucz}, R.~L. 2003, in IAU Symposium, Vol. 210, Modelling
  of Stellar Atmospheres, ed. N.~{Piskunov}, W.~W. {Weiss}, \& D.~F. {Gray}, 20

\bibitem[{{Cayrel}(1988)}]{cayrel88}
{Cayrel}, R. 1988, in IAU Symposium, Vol. 132, The Impact of Very High S/N
  Spectroscopy on Stellar Physics, ed. {G.~Cayrel de Strobel \& M.~Spite},
  345

\bibitem[{{Chiavassa} {et~al.}(2010){Chiavassa}, {Collet}, {Casagrande}, \&
  {Asplund}}]{chiavassa10}
{Chiavassa}, A., {Collet}, R., {Casagrande}, L., \& {Asplund}, M. 2010, \aap,
  524, A93

\bibitem[{{Cohen} {et~al.}(2003){Cohen}, {Wheaton}, \& {Megeath}}]{cohen03}
{Cohen}, M., {Wheaton}, W.~A., \& {Megeath}, S.~T. 2003, \aj, 126, 1090

\bibitem[{{Collet} {et~al.}(2007){Collet}, {Asplund}, \&
  {Trampedach}}]{collet07}
{Collet}, R., {Asplund}, M., \& {Trampedach}, R. 2007, \aap, 469, 687

\bibitem[{{Collet} {et~al.}(2009){Collet}, {Nordlund}, {Asplund}, {Hayek}, \&
  {Trampedach}}]{collet09}
{Collet}, R., {Nordlund}, {\AA}., {Asplund}, M., {Hayek}, W., \& {Trampedach},
  R. 2009, Memorie della Societa Astronomica Italiana, 80, 719

\bibitem[{{da Silva} {et~al.}(2006){da Silva}, {Girardi}, {Pasquini},
  {Setiawan}, {von der L{\"u}he}, {de Medeiros}, {Hatzes}, {D{\"o}llinger}, \&
  {Weiss}}]{dasilva06}
{da Silva}, L., {et~al.} 2006, \aap, 458, 609

\bibitem[{{Davis} {et~al.}(2000){Davis}, {Tango}, \& {Booth}}]{davis00}
{Davis}, J., {Tango}, W.~J., \& {Booth}, A.~J. 2000, \mnras, 318, 387

\bibitem[{{De Silva} {et~al.}(2007){De Silva}, {Freeman}, {Bland-Hawthorn},
  {Asplund}, \& {Bessell}}]{desilva07}
{De Silva}, G.~M., {Freeman}, K.~C., {Bland-Hawthorn}, J., {Asplund}, M., \&
  {Bessell}, M.~S. 2007, \aj, 133, 694

\bibitem[{{di Benedetto} \& {Foy}(1986)}]{dibenedetto86}
{di Benedetto}, G.~P., \& {Foy}, R. 1986, \aap, 166, 204

\bibitem[{{Eggen}(1971)}]{eggen71}
{Eggen}, O.~J. 1971, \pasp, 83, 271

\bibitem[{{Engelke} {et~al.}(2006){Engelke}, {Price}, \& {Kraemer}}]{engelke06}
{Engelke}, C.~W., {Price}, S.~D., \& {Kraemer}, K.~E. 2006, \aj, 132, 1445

\bibitem[{Fuhrmann}(1998)]{fuhrmann98} {Fuhrmann}, K. 1998, \aap, 338, 161

\bibitem[{{Gatewood}(2008)}]{gatewood08}
{Gatewood}, G. 2008, \aj, 136, 452

\bibitem[{{Girardi} {et~al.}(2000){Girardi}, {Bressan}, {Bertelli}, \&
  {Chiosi}}]{girardi00}
{Girardi}, L., {Bressan}, A., {Bertelli}, G., \& {Chiosi}, C. 2000, \aaps, 141,
  371

\bibitem[{{Gustafsson} {et~al.}(2008){Gustafsson}, {Edvardsson}, {Eriksson},
  {J{\o}rgensen}, {Nordlund}, \& {Plez}}]{gustafsson08}
{Gustafsson}, B., {Edvardsson}, B., {Eriksson}, K., {J{\o}rgensen}, U.~G.,
  {Nordlund}, {\AA}., \& {Plez}, B. 2008, \aap, 486, 951

\bibitem[{{Harrington} {et~al.}(1993){Harrington}, {Dahn}, {Kallarakal},
  {Guetter}, {Riepe}, {Walker}, {Pier}, {Vrba}, {Luginbuhl}, {Harris}, \&
  {Ables}}]{harrington93}
{Harrington}, R.~S., {et~al.} 1993, \aj, 105, 1571

\bibitem[{{Hinkle} \& {Wallace}(2005)}]{hinkle05}
{Hinkle}, K., \& {Wallace}, L. 2005, in Astronomical Society of the Pacific
  Conference Series, Vol. 336, Cosmic Abundances as Records of Stellar
  Evolution and Nucleosynthesis, ed. {T.~G.~Barnes III \& F.~N.~Bash}, 321

\bibitem[{{J{\o}rgensen} \& {Lindegren}(2005)}]{jorgensen05}
{J{\o}rgensen}, B.~R., \& {Lindegren}, L. 2005, \aap, 436, 127

\bibitem[{{Kiehling}(1987)}]{kiehling87}
{Kiehling}, R. 1987, \aaps, 69, 465

\bibitem[{{Kim} {et~al.}(2002){Kim}, {Demarque}, {Yi}, \& {Alexander}}]{kim02}
{Kim}, Y., {Demarque}, P., {Yi}, S.~K., \& {Alexander}, D.~R. 2002, \apjs, 143,
  499

\bibitem[{{Koesterke} {et~al.}(2008){Koesterke}, {Allende~Prieto}, \&
  {Lambert}}]{koesterke08}
{Koesterke}, L., {Allende~Prieto}, C., \& {Lambert}, D.~L. 2008, \apj, 680, 764

\bibitem[{{Kurucz} {et~al.}(1984){Kurucz}, {Furenlid}, {Brault}, \&
  {Testerman}}]{kurucz84}
{Kurucz}, R.~L., {Furenlid}, I., {Brault}, J., \& {Testerman}, L. 1984, {Solar
  flux atlas from 296 to 1300 nm} (National Solar Observatory Atlas, Sunspot,
  New Mexico: National Solar Observatory, 1984)

\bibitem[{{Lacour} {et~al.}(2008){Lacour}, {Meimon}, {Thi{\'e}baut}, {Perrin},
  {Verhoelst}, {Pedretti}, {Schuller}, {Mugnier}, {Monnier}, {Berger},
  {Haubois}, {Poncelet}, {Le Besnerais}, {Eriksson}, {Millan-Gabet}, {Ragland},
  {Lacasse}, \& {Traub}}]{lacour08}
{Lacour}, S., {et~al.} 2008, \aap, 485, 561

\bibitem[{M\"ackle} {et~al.}(1975)]{mackle75} {M\"ackle}, R., {Holweger}, H., {Griffin}, R., \& {Griffin}, R. 1975, \aap, 38, 239

\bibitem[{{Majewski} {et~al.}(2010){Majewski}, {Wilson}, {Hearty}, {Schiavon},
  \& {Skrutskie}}]{majewski10}
{Majewski}, S.~R., {Wilson}, J.~C., {Hearty}, F., {Schiavon}, R.~R., \&
  {Skrutskie}, M.~F. 2010, in IAU Symposium, Vol. 265, IAU Symposium, ed.
  {K.~Cunha, M.~Spite, \& B.~Barbuy}, 480--481

\bibitem[{{McWilliam} \& {Rich}(1994)}]{mcwilliam94}
{McWilliam}, A., \& {Rich}, R.~M. 1994, \apjs, 91, 749

\bibitem[{{Mel{\'e}ndez} {et~al.}(2009){Mel{\'e}ndez}, {Asplund}, {Gustafsson},
  \& {Yong}}]{melendez09:twins}
{Mel{\'e}ndez}, J., {Asplund}, M., {Gustafsson}, B., \& {Yong}, D. 2009, \apjl,
  704, L66

\bibitem[{{Mermilliod} {et~al.}(1997){Mermilliod}, {Mermilliod}, \&
  {Hauck}}]{mermilliod97}
{Mermilliod}, J., {Mermilliod}, M., \& {Hauck}, B. 1997, \aaps, 124, 349

\bibitem[{{Mozurkewich} {et~al.}(2003){Mozurkewich}, {Armstrong}, {Hindsley},
  {Quirrenbach}, {Hummel}, {Hutter}, {Johnston}, {Hajian}, {Elias}, {Buscher},
  \& {Simon}}]{mozurkewich03}
{Mozurkewich}, D., {et~al.} 2003, \aj, 126, 2502

\bibitem[{{Navarro} {et~al.}(2004){Navarro}, {Helmi}, \& {Freeman}}]{navarro04}
{Navarro}, J.~F., {Helmi}, A., \& {Freeman}, K.~C. 2004, \apjl, 601, L43

\bibitem[{{Nordstr{\"o}m} {et~al.}(2004){Nordstr{\"o}m}, {Mayor}, {Andersen},
  {Holmberg}, {Pont}, {J{\o}rgensen}, {Olsen}, {Udry}, \&
  {Mowlavi}}]{nordstrom04}
{Nordstr{\"o}m}, B., {et~al.} 2004, \aap, 418, 989

\bibitem[{{Perrin} {et~al.}(1998){Perrin}, {Coud{\'e} du Foresto}, {Ridgway},
  {Mariotti}, {Traub}, {Carleton}, \& {Lacasse}}]{perrin98}
{Perrin}, G., {Coud{\'e} du Foresto}, V., {Ridgway}, S.~T., {Mariotti}, J.,
  {Traub}, W.~A., {Carleton}, N.~P., \& {Lacasse}, M.~G. 1998, \aap, 331, 619

\bibitem[{{Perryman} {et~al.}(1997){Perryman}, {Lindegren}, {Kovalevsky},
  {Hoeg}, {Bastian}, {Bernacca}, {Cr{\'e}z{\'e}}, {Donati}, {Grenon}, {van
  Leeuwen}, {van der Marel}, {Mignard}, {Murray}, {Le Poole}, {Schrijver},
  {Turon}, {Arenou}, {Froeschl{\'e}}, \& {Petersen}}]{perryman97}
{Perryman}, M.~A.~C., {et~al.} 1997, \aap, 323, L49

\bibitem[{Peterson} {et~al.}(1993)]{peterson93} {Peterson}, R.~C., {Dalle 
Ore}, C.~M., \& {Kurucz}, R.~L. 1993, \apj, 404, 333 

\bibitem[{{Pont} \& {Eyer}(2004)}]{pont04}
{Pont}, F., \& {Eyer}, L. 2004, \mnras, 351, 487

\bibitem[{{Quirrenbach} {et~al.}(1996){Quirrenbach}, {Mozurkewich}, {Buscher},
  {Hummel}, \& {Armstrong}}]{quirrenbach96}
{Quirrenbach}, A., {Mozurkewich}, D., {Buscher}, D.~F., {Hummel}, C.~A., \&
  {Armstrong}, J.~T. 1996, \aap, 312, 160

\bibitem[{{Ram{\'{\i}}rez} {et~al.}(2007){Ram{\'{\i}}rez}, {Allende~Prieto}, \&
  {Lambert}}]{ramirez07}
{Ram{\'{\i}}rez}, I., {Allende~Prieto}, C., \& {Lambert}, D.~L. 2007, \aap,
  465, 271

\bibitem[{{Ram\'irez} {et~al.}(2011){Ram\'irez}, {Allende~Prieto}, {Lambert},
  \& {et al.}}]{ramirez11:thin-thick}
{Ram\'irez}, I., {Allende~Prieto}, C., {Lambert}, D.~L., \& {et al.} 2011, in
  preparation

\bibitem[{{Ram{\'{\i}}rez} {et~al.}(2006){Ram{\'{\i}}rez}, {Allende Prieto},
  {Redfield}, \& {Lambert}}]{ramirez06}
{Ram{\'{\i}}rez}, I., {Allende Prieto}, C., {Redfield}, S., \& {Lambert}, D.~L.
  2006, \aap, 459, 613

\bibitem[{{Ram{\'{\i}}rez} {et~al.}(2010){Ram{\'{\i}}rez}, {Collet}, {Lambert},
  {Allende Prieto}, \& {Asplund}}]{ramirez10:granulation}
{Ram{\'{\i}}rez}, I., {Collet}, R., {Lambert}, D.~L., {Allende Prieto}, C., \&
  {Asplund}, M. 2010, \apjl, 725, L223

\bibitem[{{Ram{\'{\i}}rez} {et~al.}(2009){Ram{\'{\i}}rez}, {Mel{\'e}ndez}, \&
  {Asplund}}]{ramirez09}
{Ram{\'{\i}}rez}, I., {Mel{\'e}ndez}, J., \& {Asplund}, M. 2009, \aap, 508, L17

\bibitem[{{Rayner} {et~al.}(2009){Rayner}, {Cushing}, \& {Vacca}}]{rayner09}
{Rayner}, J.~T., {Cushing}, M.~C., \& {Vacca}, W.~D. 2009, \apjs, 185, 289

\bibitem[{{Reddy} {et~al.}(2006){Reddy}, {Lambert}, \& {Allende
  Prieto}}]{reddy06}
{Reddy}, B.~E., {Lambert}, D.~L., \& {Allende Prieto}, C. 2006, \mnras, 367,
  1329

\bibitem[{{Reddy} {et~al.}(2003){Reddy}, {Tomkin}, {Lambert}, \&
  {Allende~Prieto}}]{reddy03}
{Reddy}, B.~E., {Tomkin}, J., {Lambert}, D.~L., \& {Allende~Prieto}, C. 2003,
  \mnras, 340, 304

\bibitem[{{Richichi} {et~al.}(2009){Richichi}, {Percheron}, \&
  {Davis}}]{richichi09}
{Richichi}, A., {Percheron}, I., \& {Davis}, J. 2009, \mnras, 399, 399

\bibitem[{Ryde} et al.(2002)]{ryde02} {Ryde}, N., {Lambert}, D.~L., {Richter}, M.~J., \& {Lacy}, J.~H. 2002, \apj, 580, 447

\bibitem[{{Schiavon} \& {Majewski}(2010)}]{schiavon10}
{Schiavon}, R.~P., \& {Majewski}, S.~R. 2010, in IAU Symposium, Vol. 262, IAU
  Symposium, ed. {G.~Bruzual \& S.~Charlot}, 428--429

\bibitem[{{Shetrone} {et~al.}(2010){Shetrone}, {Lawler}, {Schiavon},
  {Majewski}, {Hearty}, {Wilson}, {Allende Prieto}, {Johnson}, {Holtzman}, \&
  {Frinchaboy}}]{shetrone10}
{Shetrone}, M., {et~al.} 2010, in Nuclei in the Cosmos.

\bibitem[{Skrutskie} {et~al.}(2006)]{skrutskie06} {Skrutskie}, M.~F., {Cutri}, R.~M., {Stiening}, R., et~al. 2006, \aj, 131, 1163

\bibitem[{{Sneden}(1973)}]{sneden73}
{Sneden}, C.~A. 1973, PhD thesis, The University of Texas at Austin

\bibitem[{{Soubiran}(1993)}]{soubiran93}
{Soubiran}, C. 1993, \aap, 274, 181

\bibitem[{{Soubiran} {et~al.}(2010){Soubiran}, {Le Campion}, {Cayrel de
  Strobel}, \& {Caillo}}]{soubiran10}
{Soubiran}, C., {Le Campion}, J., {Cayrel de Strobel}, G., \& {Caillo}, A.
  2010, \aap, 515, 111

\bibitem[{{Tolstoy} {et~al.}(2009){Tolstoy}, {Hill}, \& {Tosi}}]{tolstoy09}
{Tolstoy}, E., {Hill}, V., \& {Tosi}, M. 2009, \araa, 47, 371

\bibitem[{Tsuji}(2009)]{tsuji09} {Tsuji}, T. 2009, \aap, 504, 543

\bibitem[{{Turner} {et~al.}(1999){Turner}, {ten Brummelaar}, \&
  {Mason}}]{turner99}
{Turner}, N.~H., {ten Brummelaar}, T.~A., \& {Mason}, B.~D. 1999, \pasp, 111,
  556

\bibitem[{{van Altena} {et~al.}(1995){van Altena}, {Lee}, \&
  {Hoffleit}}]{vanaltena95}
{van Altena}, W.~F., {Lee}, J.~T., \& {Hoffleit}, E.~D. 1995, {The general
  catalogue of trigonometric [stellar] paralaxes}, ed. {van Altena, W.~F., Lee,
  J.~T., \& Hoffleit, E.~D.}

\bibitem[{{van Leeuwen}(2007)}]{vanleeuwen07}
{van Leeuwen}, F. 2007, \aap, 474, 653

\bibitem[{{Venn} {et~al.}(2004){Venn}, {Irwin}, {Shetrone}, {Tout}, {Hill}, \&
  {Tolstoy}}]{venn04}
{Venn}, K.~A., {Irwin}, M., {Shetrone}, M.~D., {Tout}, C.~A., {Hill}, V., \&
  {Tolstoy}, E. 2004, \aj, 128, 1177

\bibitem[{{Verhoelst} {et~al.}(2005){Verhoelst}, {Bord{\'e}}, {Perrin},
  {Decin}, {Eriksson}, {Ridgway}, {Schuller}, {Traub}, {Millan-Gabet},
  {Lacasse}, \& {Waelkens}}]{verhoelst05}
{Verhoelst}, T., {et~al.} 2005, \aap, 435, 289

\bibitem[{{Williams} {et~al.}(2009){Williams}, {Freeman}, {Helmi}, \& {the RAVE
  collaboration}}]{williams09}
{Williams}, M.~E.~K., {Freeman}, K.~C., {Helmi}, A., \& {the RAVE
  collaboration}. 2009, in IAU Symposium, Vol. 254, IAU Symposium, ed.
  {J.~Andersen, J.~Bland-Hawthorn, \& B.~Nordstr{\"o}m}, 139--144

\bibitem[{{Worley} {et~al.}(2009){Worley}, {Cottrell}, {Freeman}, \& {Wylie-de-Boer}}]{worley09}
{Worley}, C.~C., {Cottrell}, P.~L., {Freeman}, K.~C., \& {Wylie-de Boer}, E.~C. 2009, \mnras, 400, 1039

\bibitem[{{Yi} {et~al.}(2001){Yi}, {Demarque}, {Kim}, {Lee}, {Ree}, {Lejeune},
  \& {Barnes}}]{yi01}
{Yi}, S., {Demarque}, P., {Kim}, Y., {Lee}, Y., {Ree}, C.~H., {Lejeune}, T., \&
  {Barnes}, S. 2001, \apjs, 136, 417

\bibitem[{{Zhang} {et~al.}(2006){Zhang}, {Butler}, {Gehren}, {Shi}, \&
  {Zhao}}]{zhang06}
{Zhang}, H.~W., {Butler}, K., {Gehren}, T., {Shi}, J.~R., \& {Zhao}, G. 2006,
  \aap, 453, 723

\bibitem[{{Zhang} {et~al.}(2008){Zhang}, {Gehren}, \& {Zhao}}]{zhang08}
{Zhang}, H.~W., {Gehren}, T., \& {Zhao}, G. 2008, \aap, 481, 489

\end{thebibliography}
\end{document}